\documentclass[a4paper,12pt]{article}
\pdfoutput=1
\usepackage{epsfig}
\usepackage{amssymb}
\usepackage{amsfonts}
\usepackage{mathtools}
\usepackage{amsmath}
\usepackage{amsthm}
\usepackage{bigints}
\usepackage{euscript}
\usepackage{verbatim}
\usepackage{latexsym}
\usepackage{graphicx}
\usepackage{caption}
\usepackage{float}
\usepackage{braket}
\usepackage{hyperref}
\hypersetup{
    colorlinks=true,
    linkcolor=black,
    citecolor=black,
    urlcolor=blue
}

\usepackage{subcaption}
\usepackage{enumitem}
\usepackage{comment}
\usepackage{tensor}
\usepackage{listings}
\usepackage{slashed}

\usepackage{mathrsfs}

\usepackage{xcolor}

\newif\ifdtup

\catcode`\@=11

\@addtoreset{equation}{section}

\def\@normalsize{\@setsize\normalsize{15pt}\xiipt\@xiipt
\abovedisplayskip 14pt plus3pt minus3pt%
\belowdisplayskip \abovedisplayskip
\abovedisplayshortskip \z@ plus3pt%
\belowdisplayshortskip 7pt plus3.5pt minus0pt}

\def\small{\@setsize\small{13.6pt}\xipt\@xipt
\abovedisplayskip 13pt plus3pt minus3pt%
\belowdisplayskip \abovedisplayskip
\abovedisplayshortskip \z@ plus3pt%
\belowdisplayshortskip 7pt plus3.5pt minus0pt
\def\@listi{\parsep 4.5pt plus 2pt minus 1pt
     \itemsep \parsep
     \topsep 9pt plus 3pt minus 3pt}}

\relax

\catcode`@=12

\catcode`\@=11

\def\section{\@startsection{section}{1}{\z@}{3.5ex plus 1ex minus
   .2ex}{2.3ex plus .2ex}{\large\bf}}

\def\SymBoxes#1#2#3#4{\newdimen\un@t \un@t#3%
\raisebox{#1}{\rule{#2\un@t}{#4}\hskip-#2\un@t
\@tempdimb\un@t \advance\@tempdimb by-#4\@tempcntb#2\relax%
\@whilenum{\@tempcntb>0}\do{
\rule{#4}{\un@t}\hskip\@tempdimb \advance\@tempcntb by\m@ne}%
\hskip-#2\un@t \rule[\un@t]{#2\un@t}{#4}%
\rule[\un@t]{#4}{#4}\hskip-#4
\rule{#4}{\un@t}}\hskip-#4}                

\begin{document}

\newcommand{\beq}{\begin{equation}}
\newcommand{\eeq}{\end{equation}}
\newcommand{\bea}{\begin{eqnarray}}
\newcommand{\eea}{\end{eqnarray}}
\newcommand{\beas}{\begin{eqnarray*}}
\newcommand{\eeas}{\end{eqnarray*}}
\newcommand{\defi}{\stackrel{\rm def}{=}}
\newcommand{\non}{\nonumber}
\newcommand{\bquo}{\begin{quote}}
\newcommand{\enqu}{\end{quote}}
\newcommand{\dd}{\mathrm{d}}
\renewcommand{\(}{\begin{equation}}
\renewcommand{\)}{\end{equation}}
\def \eqn#1#2{\begin{equation}#2\label{#1}\end{equation}}

\def\e{\epsilon}
\def\IZ{{\mathbb Z}}
\def\IR{{\mathbb R}}
\def\IC{{\mathbb C}}
\def\IQ{{\mathbb Q}}
\def\de{\partial}
\def\Tr{ \hbox{\rm Tr}}
\def\H{ \hbox{\rm H}}
\def\HE{ \hbox{$\rm H^{even}$}}
\def\HO{ \hbox{$\rm H^{odd}$}}
\def\K{ \hbox{\rm K}}
\def\Im{ \hbox{\rm Im}}
\def\Ker{ \hbox{\rm Ker}}
\def\const{\hbox {\rm const.}}
\def\o{\over}
\def\im{\hbox{\rm Im}}
\def\re{\hbox{\rm Re}}
\def\bra{\langle}\def\ket{\rangle}
\def\Arg{\hbox {\rm Arg}}
\def\Re{\hbox {\rm Re}}
\def\Im{\hbox {\rm Im}}
\def\exo{\hbox {\rm exp}}
\def\diag{\hbox{\rm diag}}
\def\longvert{{\rule[-2mm]{0.1mm}{7mm}}\,}
\def\a{\alpha}
\def\dag{{}^{\dagger}}
\def\tq{{\widetilde q}}
\def\p{{}^{\prime}}
\def\W{W}
\def\N{{\cal N}}
\def\hsp{,\hspace{.7cm}}

\def\br{\nonumber}
\def\IZ{{\mathbb Z}}
\def\IR{{\mathbb R}}
\def\IC{{\mathbb C}}
\def\IQ{{\mathbb Q}}
\def\IP{{\mathbb P}}
\def \eqn#1#2{\begin{equation}#2\label{#1}\end{equation}}

\newcommand{\C}{\ensuremath{\mathbb C}}
\newcommand{\Z}{\ensuremath{\mathbb Z}}
\newcommand{\R}{\ensuremath{\mathbb R}}
\newcommand{\rp}{\ensuremath{\mathbb {RP}}}
\newcommand{\cp}{\ensuremath{\mathbb {CP}}}
\newcommand{\vac}{\ensuremath{|0\rangle}}
\newcommand{\vact}{\ensuremath{|00\rangle}                    }
\newcommand{\oc}{\ensuremath{\overline{c}}}
\newcommand{\psizero}{\psi_{0}}
\newcommand{\phizero}{\phi_{0}}
\newcommand{\hzero}{h_{0}}
\newcommand{\psiin}{\psi_{\rh}}
\newcommand{\phiin}{\phi_{\rh}}
\newcommand{\hin}{h_{\rh}}
\newcommand{\rh}{r_{h}}
\newcommand{\rb}{r_{b}}
\newcommand{\psibnd}{\psi_{0}^{b}}
\newcommand{\psibndp}{\psi_{1}^{b}}
\newcommand{\phibnd}{\phi_{0}^{b}}
\newcommand{\phibndp}{\phi_{1}^{b}}
\newcommand{\gbnd}{g_{0}^{b}}
\newcommand{\hbnd}{h_{0}^{b}}
\newcommand{\zh}{z_{h}}
\newcommand{\zb}{z_{b}}
\newcommand{\man}{\mathcal{M}}
\newcommand{\hbr}{\bar{h}}
\newcommand{\tbr}{\bar{t}}
\newcommand{\zbr}{\bar{z}}
\newcommand{\wbr}{\bar{w}}

\newcommand{\tlam}{\tilde\lambda}
\newcommand{\tOm}{\tilde\Omega}
\newcommand{\tC}{\tilde{\mathcal{C}}}
\newcommand{\tD}{\tilde{\mathcal{D}}}
\newcommand{\tE}{\tilde{\mathcal{E}}}
\newcommand{\tal}{\tilde\alpha}

\newcommand{\scrip}{\mathscr{I}^{+}}
\newcommand{\scrim}{\mathscr{I}^{-}}
\newcommand{\scri}{\mathscr{I}}

\begin{titlepage}
\begin{flushright}
\end{flushright}
\bigskip
\def\thefootnote{\fnsymbol{footnote}}

\begin{center}
{\large
{\bf Holography, Brick Wall and a Little Hierarchy Problem
}
}
\end{center}

\bigskip
\begin{center}
Vishal Gayari$^a$\footnote{\texttt{vishalgayari@iisc.ac.in}}, Chethan Krishnan$^a$\footnote{\texttt{chethan.krishnan@gmail.com}}, \ Pradipta S. Pathak$^a$ \footnote{\texttt{pradiptap@iisc.ac.in}}
\vspace{0.1in}

\end{center}

\renewcommand{\thefootnote}{\arabic{footnote}}

\begin{center}
\vspace{-0.2cm}

$^a$ {\it Center for High Energy Physics, Indian Institute of Science, \\ C V Raman Road, Bangalore 560012, India}\\

\end{center}

\noindent
\begin{center} {\bf Abstract} \end{center}
We propose a heuristic for the brick wall in AdS/CFT: the location where a boundary mode’s local bulk energy reaches a (Planckian) UV cut-off. This accomplishes two things: (a) the brick wall is framed as a breakdown criterion for bulk effective field theory, and (b) the definition is boundary-anchored rather than horizon-anchored, aligning it with holography. Near the horizon, spacetime effectively gets cut-off due to blueshift relative to the boundary, and leads to normal modes. By directly computing these new modes for the BTZ black hole, we show that they are qualitatively unchanged from conventional 't Hooftian brick wall normal modes in the relevant part of the spectrum -- successfully reproducing black hole thermodynamics and exterior smooth-horizon correlators, under similar approximations. However, unlike 't Hooft's (and our own previous) calculations, we also do an {\em exact} numerical evaluation of the normal mode partition function. This allows us to identify a ``little hierarchy" problem in the brick wall paradigm, irrespective of whether it is horizon-anchored or boundary-anchored: because the modes are not exactly degenerate in the $J$-direction, the coefficient of the area law is slightly subleading, unless the brick wall is slightly trans-Planckian. One way to evade the problem is to increase the number of active species. While this is certainly a possibility in string theory, we argue that a natural resolution is to take into account the degrees of freedom {\em intrinsic} to the (stretched) horizon, as suggested by the recent results in \cite{CKPSP}. We argue that this will lead to a dominant contribution from a quantum number associated to the {\em radial} direction, while retaining the successes of the $J$-degenerate toy model. We discuss the possible significance of these observations for (a) quantum chaos in black holes, and (b) the fuzzball program.


\vspace{1.6 cm}
\vfill

\end{titlepage}

\setcounter{footnote}{0}


\tableofcontents

\section{Brick Walls}

In recent papers, we have argued that the brick wall is a useful ingredient for building toy models for black hole microstates \cite{Burman1, Burman2}: see also closely related work in \cite{Adepu}. The brick wall as introduced by 't Hooft \cite{tHooft} is a Dirichlet boundary condition for fields outside the event horizon. We will work with a single scalar field in this paper, but generalization to more (free) fields is straightforward. In the following we will often consider the BTZ black hole and use notation adapted to it, but the observations generalize to more general black holes \cite{Pradipta-holomorphic}. 

The main point of \cite{Burman1, Burman2} was that quantum fields with such a brick wall boundary condition can be used to motivate toy model spectra that precisely reproduce the thermodynamics and exterior smooth horizon correlators\footnote{More precisely, the correlators in the exterior region of the black hole are indistinguishable from smooth horizon correlators up to corrections that are exponentially suppressed in the entropy, till the Page time. This matches expectations from eigenstate thermalization \cite{Deutsch}.} of black holes, while working within a non-lossy setting. In \cite{Burman1, Burman2}, the following observations were emphasized as key features of a brick wall(-inspired) spectrum:
\begin{itemize}
\item  The spectrum of normal modes is quasi-degenerate in the angular quantum number $J$. This is the underlying origin of the area law for entropy\footnote{The normal mode spectrum has a weak dependence on the $J$-quantum number because gravitational redshift near the horizon effectively suppresses the growth of modes along one of the spatial directions. This is the dynamical origin of the area law in brickwall models. Importantly, area scaling is {\em not} a result of any cutoff along $J$-direction, i.e. $J_{\rm cut}$ or $J_{\rm sat}$ (later in the text): it is the {\em coefficient} of the area law that these quantities determine. In comparison, Planckian black body obeys volume law because the modes can grow linearly along all the spatial directions.}. (See also \cite{tHooft, Pradipta1}.)
\item  The exact spectrum --in particular, its precise (weak) $J$-dependence-- is only known numerically. This makes working with the exact partition function, difficult. On the other hand, if one simply {\em assumes} exact degeneracy in $J$, the partition function (naturally) diverges, and therefore an explicit $J_{\rm cut}$ is needed. This $J_{\rm cut}$ was fixed in \cite{Burman1, Burman2} by the demand that the coefficient of Bekenstein-Hawking entropy comes out right. This completely fixes the normal mode spectrum and the calculations, successfully. 
\end{itemize}

\noindent
[In Appendix \ref{sec:Jcut} of this paper, we will note that the $J_{\rm cut}$ can also be understood in a different way. It can also be fixed by the demand that we only retain the (degenerate-in-$J$) normal modes that satisfy a basic requirement: that at a given $J$ they are  not {\em below} the potential\footnote{This observation was first made by one of us (VG) in unpublished work with Suman Das and CK.}. This connects with 't Hooft's own approach, which adopted a semi-classical Sommerfeld quantization path to obtain the area law for normal modes. We will explain later in Appendix \ref{sec:tHooft} that the real utility of 't Hooft's semi-classicality demand is that it makes the resulting {\em approximate} normal modes, more degenerate than the {\em actual} normal modes, aligning it with our two bullet points above. In particular, the semi-classical normal modes go ``under" the potential just as the degenerate modes do, naturally providing a cut-off in $J$.]

One lesson from \cite{Burman1, Burman2} was that degenerate normal modes together with a cut-off in $J$ was sufficient to understand black hole thermodynamics and two-point correlators up to the Page time, in a unitary setting\footnote{Note that a smooth horizon results in quasinormal modes, which cause information loss. The results in \cite{Burman1, Burman2} connect with this fact in the vanishing limit of the stretched horizon: the exact correlator reduces to the Hartle-Hawking correlator.}. Remarkably, the precise location of the brick wall dropped out of the calculation, as long as it was close enough to the horizon to justify the quasi-degeneracy approximation. 

In this paper, we will compute the exact partition function of  the brick wall normal modes numerically. This allows us to get a sense of {\em how} drastic an approximation is being made when we take the spectrum to be degenerate-in-$J$, together with the $J_{\rm cut}$. A crucial fact that enables this calculation is the {\em finiteness} of the partition function when we work with the exact spectrum. This is a conceptual improvement over the degenerate approximation, where the cut-off in $J$ was a kinematical input. (Note that without a $J_{\rm cut}$, a degenerate spectrum will lead to an arbitrary number of $J$-modes being populated without any energetic cost.) The exact calculation bypasses this, because even though the $J$-dependence is weak, it is sufficient to saturate the partition function without the explicit introduction of a $J_{\rm cut}$. The direct calculation also gives us an opportunity to critically evaluate the brick wall as a model for black hole microstates: this is possible because the results and conclusions of the exact calculation are qualitatively quite robust.

A main conclusion of the exact calculation is that while the quasi-degeneracy of the spectrum does lead to an area-scaling, a precise matching requires the brick wall to be hierarchically closer to the horizon than the Planck length. In our numerical results for the BTZ black hole, this hierarchy is about $\sim \mathcal{O}(10^{-3})$. While this may not seem like a particularly bad hierarchy, we find that it is quite persistent as we change the parameters of the black hole. Furthermore, it is present in other black holes (for Schwarzschild, it is $\sim10^{-4}$-$10^{-5}$) as well\footnote{We will show in the Appendix~\ref{sec:tHooft}, that under the hood, there is indeed a little hierarchy in 't Hooft's old calculation \cite{tHooft} of Schwarzschild as well. In work that we will not include here, we have also checked similar hierarchies in some other black holes. While the precise hierarchy varies, its existence seems generic.}. We also find that the coefficient of the energy and entropy do not precisely match, even if we accept the hierarchy: there is an $\mathcal{O}(1)$ difference from black hole values in the energy/mass, if we decide to accept the hierarchy in a way that matches the coefficient of the Bekenstein-Hawking entropy. All these observations are consistent with the expectation that the degeneracy is a key input in getting the area law, and that breaking the degeneracy (even relatively weakly) is analogous to ``weakly turning on an extra dimension".

While there are a few ways to evade the little hierarchy problem\footnote{Most notably, by allowing about $\sim 100$ - $1000$ species of particles at the Planck scale (this specific count refers to BTZ). Note that including spin/polarization degrees of freedom, this is far from an exorbitant number in typical BSM models.}, in this paper, we will adopt the point of view that it is an indication that a toy probe field can only go so far: we need to take into account the degrees of freedom intrinsic to the (stretched) horizon. This is natural from the perspective of some recent developments \cite{CKPSP}, where it was observed that the partition function of a smooth horizon Euclidean BTZ black hole has a re-writing as a {\em sum} over states without a smooth horizon. Notably, the sum was over states with a $U(1)$ isometry, indicating that the dominant degeneracy should arise from a suitably defined radial quantum number. We will make some comments about the implications of this observation for the fuzzball program in the concluding section.

But before turning to the above-mentioned exact calculation of the partition function, we will formulate the problem in a holography-compatible way. The 't Hooftian brick wall is usually defined \cite{tHooft, Susskind} with respect to the horizon: it is taken to be some physical distance (Planck/string scale) away from the actual horizon. This is an ad-hoc cut-off in spacetime. We would like to improve this in two ways. Firstly, we wish to have a definition of the brick wall that is holography-compatible: we would like to define its location with respect to the boundary, and not the horizon. Secondly, instead of postulating it as an ad-hoc location, we would like it to emerge as a location in the bulk where bulk effective field theory breaks down. The goal of these refinements is to allow the brick wall paradigm to put its best foot forward.

It turns out that both these improvements can be accomplished in the following way: we demand that a mode corresponding to the boundary frequency $\omega$, vanishes at the bulk location where its local bulk energy reaches the bulk UV cut-off -- which we will take to be a scale comparable to the Planck scale. It turns out this condition can be used to give an alternate definition of the normal modes, and they are qualitatively unchanged (in the regime that is relevant for black hole thermodynamics) from the 't Hooftian normal modes\footnote{Note that in the conventional brick wall, even though it is viewed heuristically as a bulk UV cut-off, some modes reach trans-Planckian bulk energies at some places in the bulk. The holography-compatible definition does not have this problem, but because of the mode-dependence of the cut-off, it is not a conventional Sturm-Liuoville problem. Our eventual attitude will be that neither of these approaches are fully satisfactory.}. We will mostly present our exact partition function calculation with these holographic normal modes, but we have checked that the conclusions are unaffected if we work with the exact 't Hooftian normal modes as well. We will discuss the features of the spectrum that are responsible for this robustness in due course -- essentially it arises directly from the non-constancy (more precisely, the weakly logarithmic form) of the $J$-dependence. As may be suspected, we find that the holography compatible modes do marginally {\em better} than the conventional brick wall normal modes: but both suffer from the little hierarchy.

The main goal of this paper was to identify the {\em crudest} aspect of the brick wall approximation, so that one can try to understand how to go beyond it. The 't Hooftian brick wall \cite{tHooft} crucially relies on an approximate degeneracy-in-$J$ to produce its area scaling \cite{tHooft}. The works of \cite{Burman1, Burman2} used this as a hint to construct a toy spectrum that could reproduce the precise {\em quantitative} features of black hole thermodynamics and correlators\footnote{We expect the main claims of \cite{Burman1, Burman2} to remain valid for {\em any} mechanism that can produce an exponential density of states consistent with the precise Bekenstein-Hawking area law for black hole entropy \cite{VishalVaibhav}.}. In this paper, we will show that moving away from the toy spectrum and working exactly in the brick wall normal modes leads to a little hierarchy problem. We will argue that this arises because the brick wall misses degrees of freedom that are more naturally viewed as intrinsic to the (stretched) horizon. Part of the motivation for this, came from the results of  \cite{CKPSP}. A more realistic model that incorporates these degrees of freedom seems possible in light of \cite{CKPSP}. We hope to report on it in upcoming work.

\section{Mode Spectrum}\label{sec:BTZ}

We will mostly work with the non-rotating BTZ black hole whose exterior geometry is described by
\begin{align} \label{sphermetric}
\dd{s}^2 = -f(r)\dd{t}^2 + \frac{\dd{r}^2}{f(r)} + r^2 \dd{\phi}^2 \hspace{1cm} \, \quad \quad {\rm where} \ f(r) = \frac{r^2-r^{2}_{h}}{L^2} 
\end{align}
with $-\infty < t < \infty$, $r_h < r < \infty$ and $0 \leq \phi < 2\pi$. For a probe scalar of mass $m$, after choosing the ansatz 
\beq \label{BTZansatz}
\Phi(t,r,\phi) = \sum_{\omega,J} e^{-i\omega t}e^{i J \phi} R_{\omega,J}(r)
\eeq  
the radial differential equation in this background is:
\beq \label{radialdiff}
f(r)R^{''}_{\omega,J}(r) + \left(\frac{r f^{'}(r) + f(r)}{r}\right)R^{'}_{\omega,J}(r)+\left(\frac{\omega^2}{f(r)} - \frac{J^2}{r^2} - \frac{\Delta (\Delta - 2)}{L^2}\right)R_{\omega,J}(r) = 0
\eeq
which has the normalizable mode solution \cite{Pradipta1} : 
\beq \label{BTZnormsoln}
R_{\omega,J}(r) = C_1 \left(\frac{r_h}{r}\right)^{\Delta} \left(1 - \frac{r^2_h}{r^2}\right)^{c}{}_{2}F_{1}\left(a, b, \Delta, \frac{r^2_h}{r^2}\right)
\eeq
where $C_1$ is an $r$-independent constant and $a,b,c$ are given by 
\begin{align}
a = \frac{\Delta}{2} + \frac{i L (J - \omega L)}{2 r_h}, \, \quad b = \frac{\Delta}{2} - \frac{i L (J + \omega L)}{2 r_h}, \, \quad \ c = -\frac{i \omega L^2}{2 r_h}
\end{align}
and $\Delta$ is the scaling dimension of the probe in dual CFT, i.e., $\Delta = 1 + \sqrt{1 + m^2 L^2}$. 

We will define a set of ``stretched horizon" normal modes in this paper that are (a) natural from the boundary point of view, and (b) incorporates the idea that the stretched horizon is a bulk UV cut-off. In other words, we will provide a boundary-anchored (or holographic) definition of the stretched horizon/brick wall. Using the standard prescription for relating proper time in the bulk ($\tau$) to the boundary time ($t$) using \eqref{sphermetric},
\beq \label{bndbulkomega}
f(r)dt^2 = d\tau^2 \implies \frac{\omega}{\sqrt{f(r)}} = \omega_{\rm bulk}
\eeq
we define the stretched horizon as the location ($r_\epsilon = r_h + \epsilon$) where the local bulk energy $\omega_{\rm bulk}$ gets blue-shifted relative to the boundary mode $\omega$, and reaches a bulk UV cut-off:
\beq \label{boundarydef}
\frac{\omega}{\sqrt{f(r_\epsilon)}} \equiv M = \frac{1}{\alpha \ell_p} \implies r_\epsilon \equiv r_h \sqrt{1+\left(\frac{\omega L \alpha \ell_p}{r_h}\right)^2}
\eeq
where $M$ is the bulk UV cut off and $\ell_p$ is the Planck length. The factor $\alpha$ controls the hierarchy between the two. Note that this holography-inspired definition\footnote{We could try a slightly more general version of bulk UV cut-off, by allowing rotational mode energy also to contribute: 
\bea
\frac{\omega^2}{{f(r)}}-\frac{J^2}{r^2} = \omega^2_{\rm bulk}.
\eea
But we find that our conclusions are unaffected by such a choice, so we will stick with the simpler definition.} of the cut-off treats $\epsilon$ as a function of $\omega$, whereas earlier works used a hard cutoff for all $\omega$.

For comparison with the traditional hard brick-wall construction \cite{tHooft, Burman1}, we will also state the definition of the latter. In this approach, the cut-off is placed at a fixed geodesic distance ($s$), taken to be of the order of Planck length, from the physical horizon:
\beq \label{slp}
s \approx L\sqrt{\frac{2 \epsilon}{r_h}} \equiv \alpha \ell_p = \frac{1}{M}\implies r_\epsilon \equiv r_h \left(1 + \frac{\alpha^2 \ell_p^2}{2L^2} \right)
\eeq
Note that this cut-off is $\omega$-$in$dependent.

We can now solve \eqref{BTZnormsoln} for the boundary mode spectrum ($\omega$), after imposing that the mode vanishes at the stretched horizon. The problem gets numerically challenging mainly for two reasons: (a) at a fixed $\ell_p$, the magnitude of the function grows rapidly as $J$ increases, leading to numerical instability in software like Mathematica or Python, (b) decreasing $\ell_p$ leads to the stretched horizon approaching the event horizon, where the hypergeometric function approaches a branch cut. For instance, when $r_h/L = 1$, the closeness to the branch cut destabilizes the numerics around $\ell_p/L \sim  10^{-8}$. 

\begin{figure}
\centering
\begin{subfigure}{.5\textwidth}
  \centering
  \includegraphics[width=0.9\linewidth]{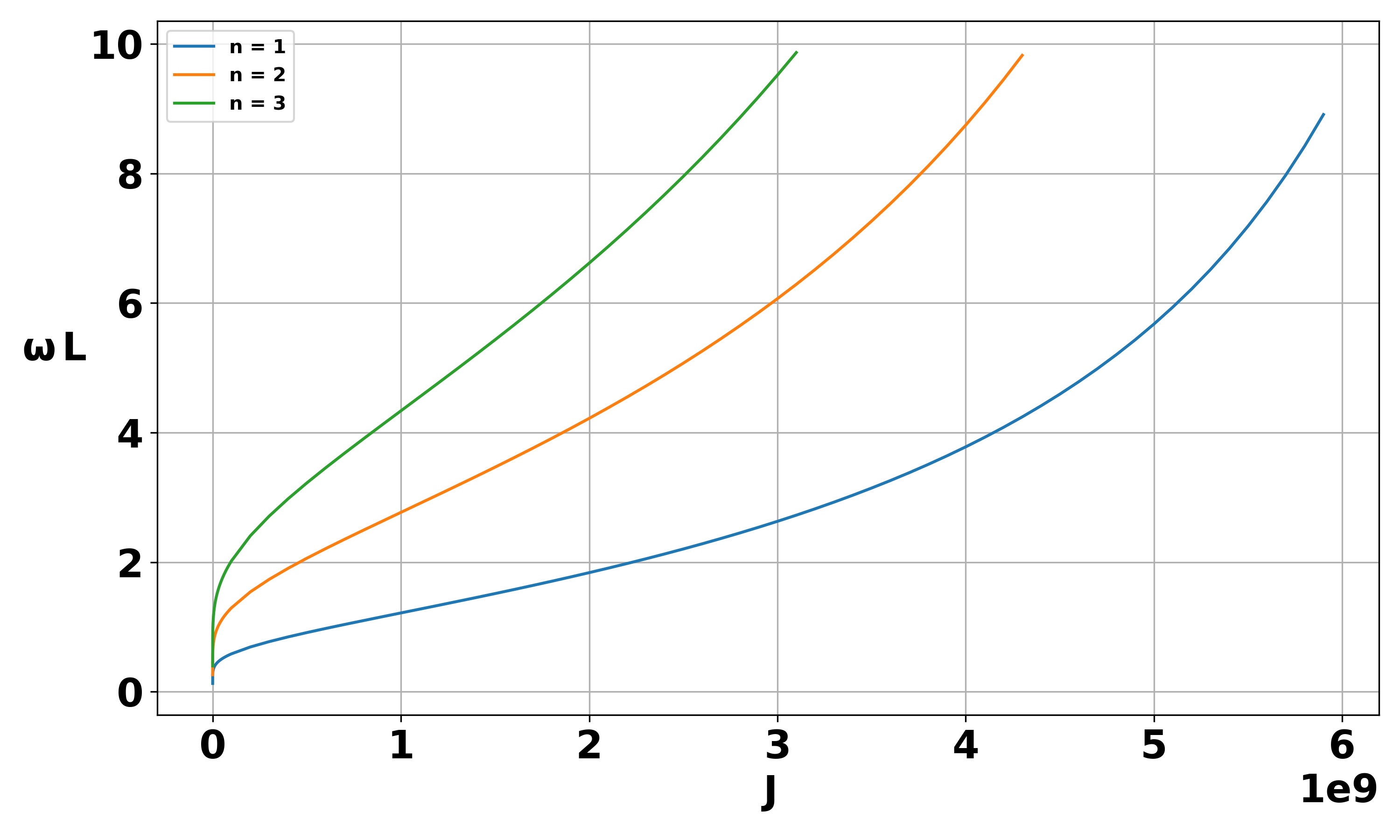}
  \caption{}
  \label{HSHspectrum}
\end{subfigure}%
\begin{subfigure}{.5\textwidth}
  \centering
  \includegraphics[width=0.9\linewidth]{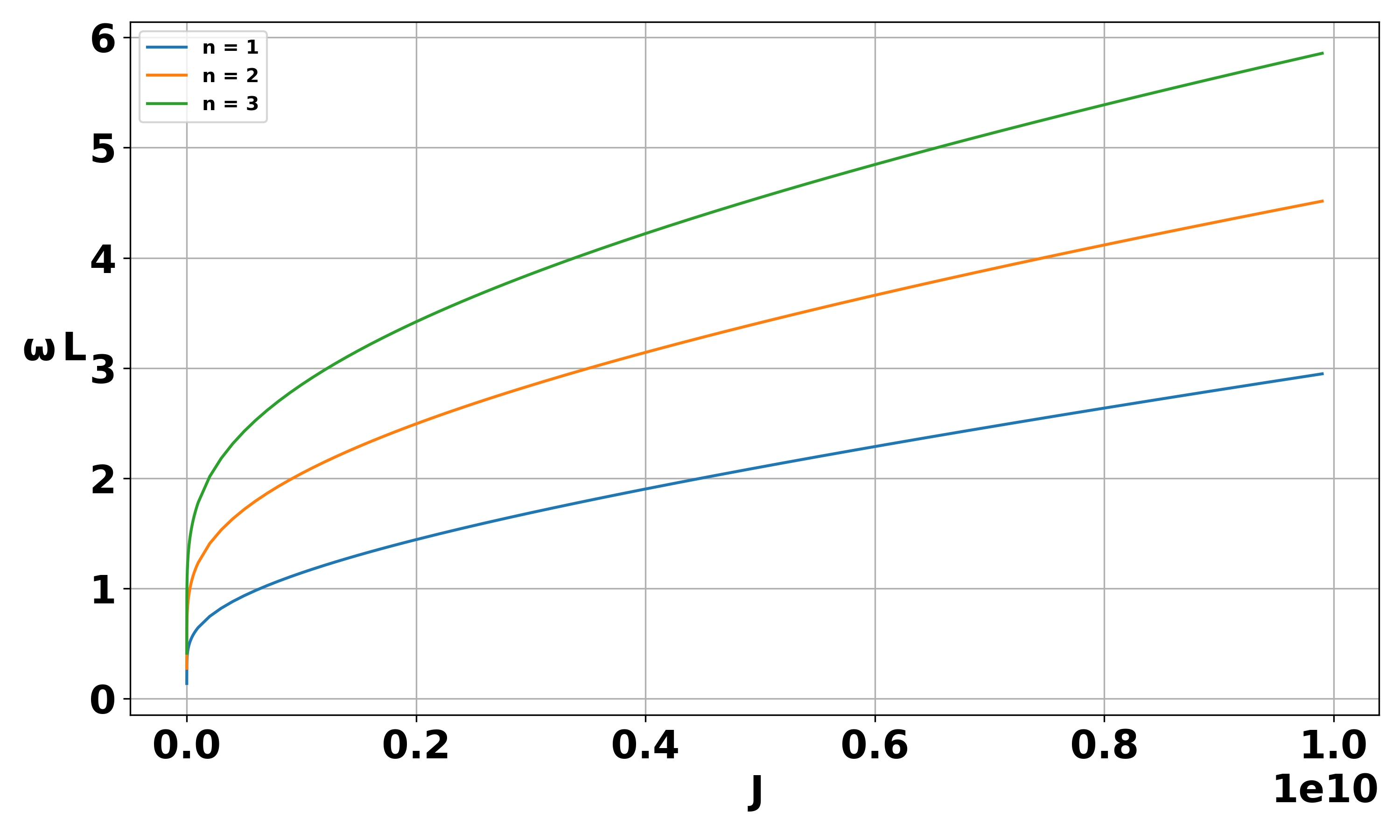}
  \caption{}
  \label{}
\end{subfigure}
\caption{Spectrum from BTZ Exact Phase Equation in (a) the boundary-anchored and (b) the hard-wall cutoff case with $\widetilde \ell = 10^{-10}$ and $r_h/L = 1$ for $n=1,2$ and $3$.}
\label{bndancandhard}
\end{figure}

To make things numerically tractable, we first convert \eqref{BTZnormsoln} using the Kummer relation
\bea \label{Kummer}
{}_2F_1(a,b,\Delta,z) &=& 
\frac{\Gamma(\Delta)\Gamma(a + b - \Delta)}{\Gamma(a)\Gamma(b)} (1 - z)^{\Delta - a - b}
{}_2F_1(\Delta - a, \Delta - b, \Delta - a - b + 1, 1 - z) \nonumber \\
&+& \frac{\Gamma(\Delta)\Gamma(\Delta-a-b)}{\Gamma(\Delta-a)\Gamma(\Delta-b)}
{}_2F_1(a, b, a + b - \Delta + 1, 1 - z) 
\eea
and then impose vanishing boundary condition at $r_\epsilon$ to get:
\beq \label{BTZexact}
\left(\frac{r_h}{r_\epsilon}\right)^{\Delta} {\rm Re}\left[\left(1 - \frac{r^2_h}{r^2_\epsilon}\right)^{c} \ \frac{\Gamma(2 c^*)}{\Gamma(a^*)\Gamma(b^*)} \ {}_{2}F_{1}\left(a, b, 1 + 2 c, 1 - \frac{r^2_h}{r^2_\epsilon}\right)\right] = 0
\eeq
where $\rm Re$ stands for the Real part and $*$ denotes complex conjugation. The $r_\epsilon$ here can be the conventional stretched horizon or the $\omega$-dependent holographic one. The LHS can be re-written using the relation $\rm Re[z] = |z| \cos(Arg[z]), \ \forall \ z \in \mathbb C$: 
\beq \label{BTZexactphase}
{\rm Cos}\left({\rm Arg}\left[{}_{2}F_{1}\left(a, b, 1 + 2 c, 1 - \frac{r^2_h}{r^2_\epsilon}\right)\right] + {\rm Arg}\left[\frac{\Gamma(a) \Gamma(b)}{\Gamma(2 c)}\right] - \frac{\omega L^2}{2 r_h} \log \left(1 - \frac{r^2_h}{r^2_\epsilon}\right)\right) = 0
\eeq
upto an overall prefactor which does not contribute to the roots. We call this the {\em BTZ Exact Phase Equation} and solve it  numerically for the spectrum $(\omega L)$ of a massless probe ($\Delta = 2$), with a choice of $r_h/L$ and a dimensionless UV cut-off
\beq \label{lptilde}
\widetilde \ell \equiv\frac{1}{M L} = \frac{\alpha \ell_p}{L} 
\eeq
for both boundary anchored \eqref{boundarydef} and hard wall cutoff \eqref{slp} case, as shown in Fig. \ref{bndancandhard}. 

\begin{figure}
\centering
\begin{subfigure}{.5\textwidth}
  \centering
  \includegraphics[width=0.9\linewidth]{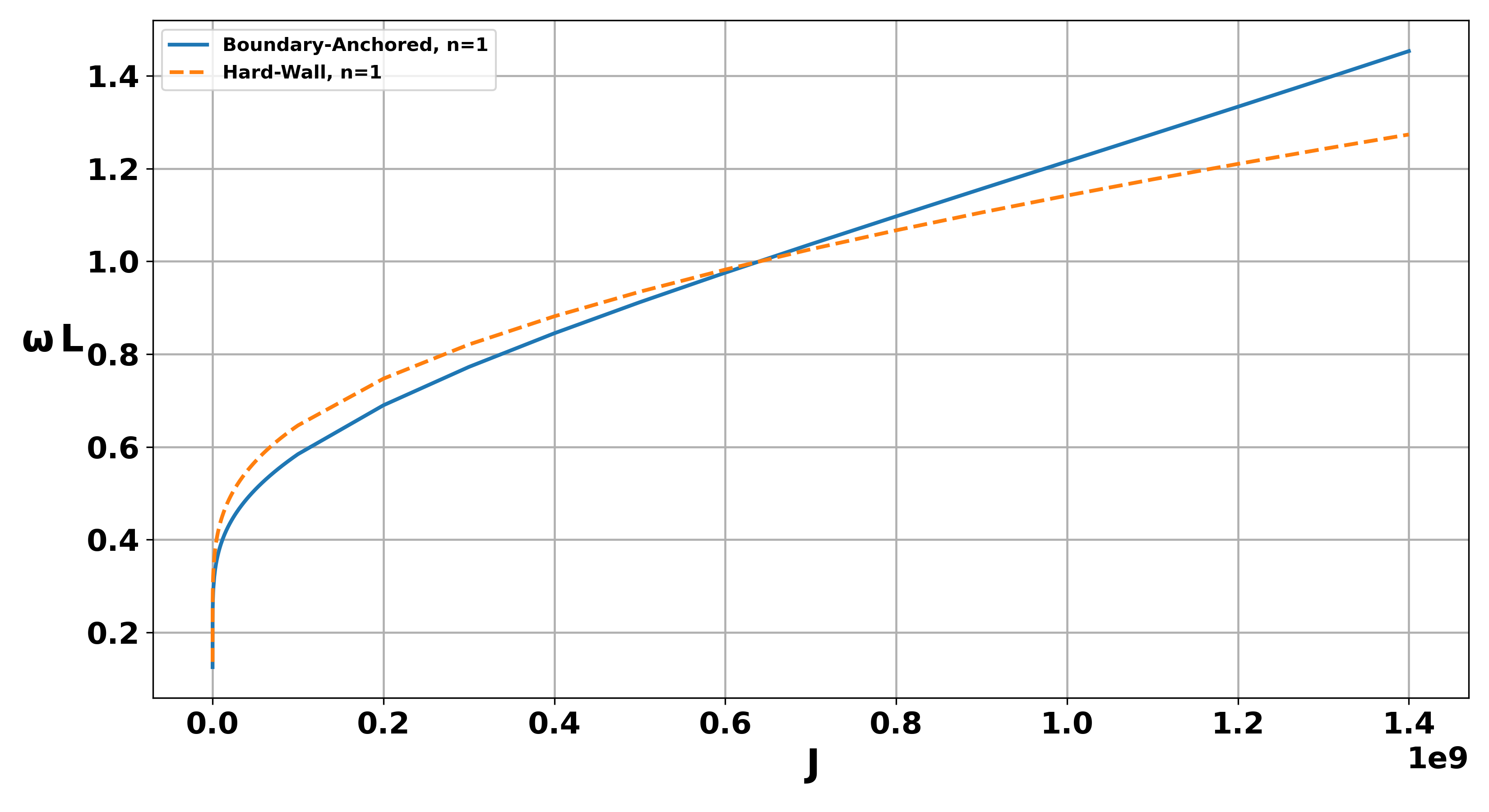}
  \caption{}
  \label{}
\end{subfigure}%
\begin{subfigure}{.5\textwidth}
  \centering
  \includegraphics[width=0.9\linewidth]{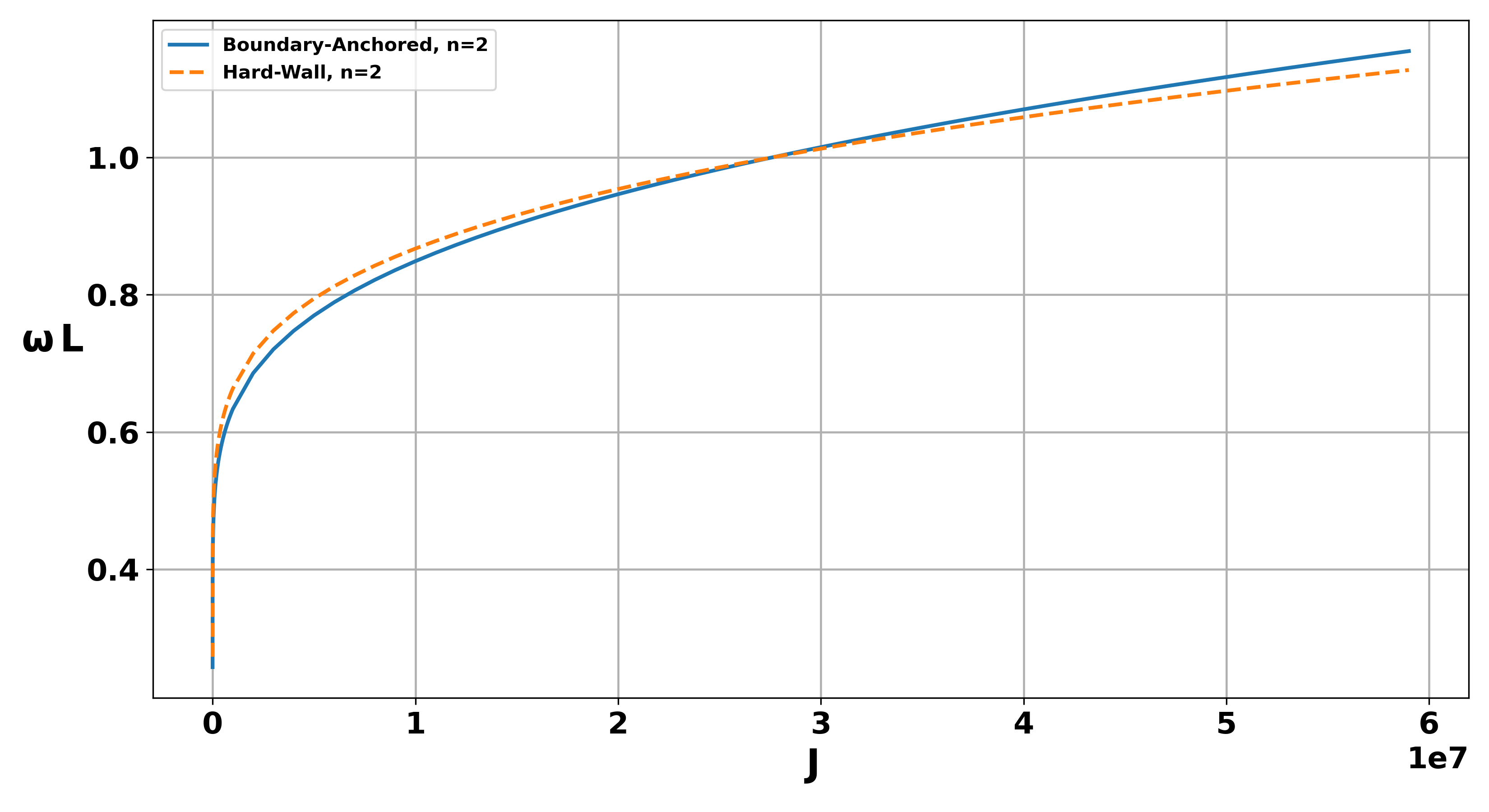}
  \caption{}
  \label{}
\end{subfigure}
\caption{Comparison between the exact boundary-anchored spectrum (Blue) and the hardwall spectrum (yellow) with $\widetilde \ell = 10^{-10}$ and $r_h/L = 1$, for (a) $n=1$ and (b) $n=2$.}
\label{bdryancandhoranc}
\end{figure}

Once we have both boundary anchored and hard wall spectra, the next step is to compare them, which we do in Fig. \ref{bdryancandhoranc}. It has been shown previously \cite{Pradipta1, Burman1} that thermodynamics derived from the normal modes receives a dominant contribution from the low-lying part of this spectrum. In the plot, this region corresponds to frequencies up to the crossover point at $\omega L \sim 1$, in which the boundary-anchored spectrum lies below the hard-wall spectrum. This relative downward shift implies a greater degeneracy (in $J$) of the holographic normal modes, in the thermodynamically relevant region, than in the hard-wall normal modes.

It is instructive to find an analytical approximate expression for this region. We can make progress by focusing on the low-energy sector, i.e. $\omega L \ll 1$. In this limit, the exact spectrum computed from \eqref{BTZexactphase} is not affected by the presence of the hypergeometric term. This can be seen from Fig. \ref{HSHspectrum} where the solution with and without the hypergeometric term are plotted: for $\omega L \ll 1$, they match. The remaining terms admit a systematic expansion\footnote{A closely related approximation was reported earlier (see Eqn (5.12) and (5.13) of \cite{Pradipta1}), albeit obtained via numerical curve fitting. Here we show that it can, in fact, be obtained completely analytically by using the identity, ${\rm Arg}[z] = {\rm Im}[\log(z)]$ where ${\rm Im}$ is the imaginary part.}:
\bea \label{appxloww}
\begin{aligned}
{\rm Arg}\!\left[\frac{\Gamma(a)\Gamma(b)}{\Gamma(2c)}\right]
&\approx -\frac{\pi}{2}
-\frac{\omega L^2}{2 r_h} 
\left(
\psi\!\left(\frac{\Delta}{2}-\frac{iJL}{2r_h}\right)
+\psi\!\left(\frac{\Delta}{2}+\frac{iJL}{2r_h}\right)
+2\gamma
\right)
+\mathcal{O}(\omega^3 L^3)
\\[8pt]
\frac{\omega L^2}{2 r_h}
\log\!\left(1-\frac{r_h^2}{r_\epsilon^2}\right)
&\approx
\begin{cases}
\dfrac{\omega L^2}{r_h}
\log\!\left(\dfrac{\omega L^2 \widetilde\ell}{r_h}\right)
+ \mathcal{O}(\omega^2 L^2),
& \text{boundary anchored} . \\[12pt]
\dfrac{\omega L^2}{r_h}
\log\! \ \widetilde\ell
+ \mathcal{O}\left(\widetilde \ell^2\right), 
& \text{hard wall} .
\end{cases}
\end{aligned}
\eea
where the hard wall case is an expansion in $\widetilde \ell$, $\gamma = 0.5772...$ is the Euler-Mascheroni constant and $\psi(z)$ is the Digamma function.

\begin{figure}
\centering
\begin{subfigure}{.5\textwidth}
  \centering
  \includegraphics[width=0.9\linewidth]{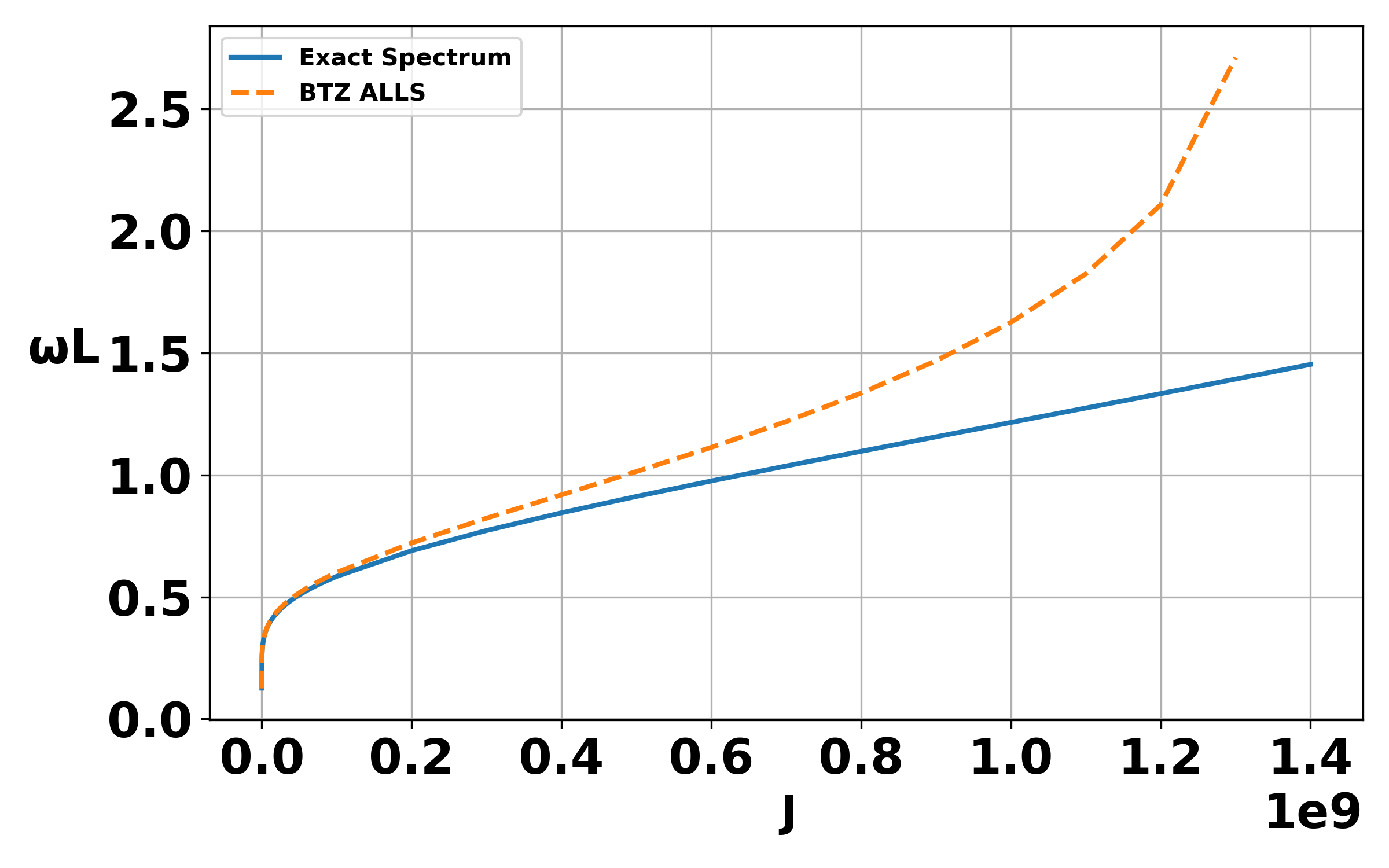}
  \caption{}
  \label{HSHspectrum}
\end{subfigure}%
\begin{subfigure}{.5\textwidth}
  \centering
  \includegraphics[width=0.9\linewidth]{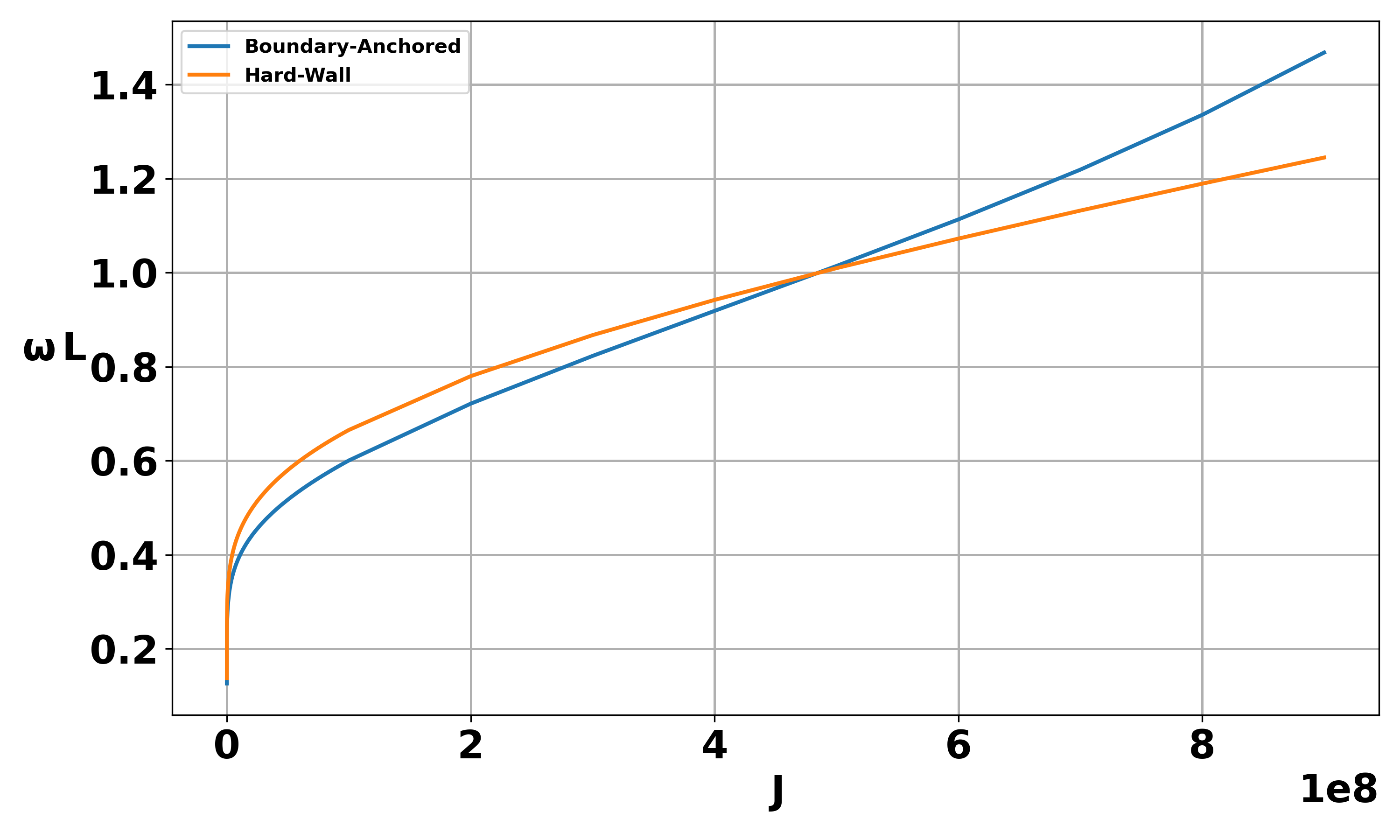}
  \caption{}
  \label{bdryvshardALLS}
\end{subfigure}
\caption{(a) Exact spectrum (blue) vs holographic {\em ALLS} (yellow). (b) Holographic {\em ALLS} in \eqref{BTZALLS} (blue) and Hardwall {\em ALLS} \eqref{BTZhardwall} (yellow) with $\widetilde \ell = 10^{-10}$ and $r_h/L = 1$, for $n = 1$.}
\label{}
\end{figure}

Using the fact that ${\rm cos}(z) = 0$ when $z = (2n-1)\pi/2$, we can write an analytical expression satisfied by the boundary anchored case, which is valid for low $\omega L$:
\beq \label{BTZALLS}
\frac{\omega L^2}{r_h}\approx \frac{2 n\pi}{2\log\left(\frac{L}{\alpha l_{p}}\right) - \left(\psi\left(\frac{\Delta}{2} - \frac{i J L}{2 r_h}\right) + \psi\left(\frac{\Delta}{2} + \frac{i J L}{2 r_h}\right) + 2 \gamma\right) - 2\log \left(\frac{\omega L^2}{r_h}\right)},
\eeq
where $n \in \mathbb Z^+$. We call this the {\em BTZ Analytical Low-Lying Spectrum (ALLS)} for the holographically anchored case. An analogous expression for the hard-wall case was reported in eqn. $(2.18)$ of \cite{Burman1}\footnote{For a massless probe, eqn. $(2.18)$ of \cite{Burman1} is a good approximation to the general expression \eqref{BTZhardwall} that we present here, from $JL/r_h \sim 2$ to arbitrarily high values of $JL/r_h $. This statement is true up to a small number ($2 \gamma - \log 4 = -0.2318...$ to be precise) which can be neglected for small enough $\widetilde \ell$.}. We will write this below in a slightly more general form coming from \eqref{appxloww}: 
\beq \label{BTZhardwall}
\frac{\omega L^2}{r_h} \approx \frac{2 n\pi}{2\log\left(\frac{L}{\alpha \ell_{p}}\right) - \left(\psi\left(\frac{\Delta}{2} - \frac{i J L}{2 r_h}\right) + \psi\left(\frac{\Delta}{2} + \frac{i J L}{2 r_h}\right) + 2 \gamma\right)},
\eeq
The thing to note is that the only difference between \eqref{BTZALLS} vs \eqref{BTZhardwall} is the presence of an extra term, $\log \left(\frac{\omega L^2}{r_h}\right)$, in the denominator. However, the thermodynamically relevant regime ($\frac{\omega L^2}{r_h} \sim \mathcal{O}(1)$, shown in Section~\ref{bulkplankian}) of the spectrum is qualitatively the same with/without this term, as shown in Fig. \ref{bdryvshardALLS}. A direct consequence of this is that the spectrum is still approximately $J$-degenerate. Therefore, the approximation of exact $J$-degeneracy can be motivated from both spectra:
\beq \label{BTZdegenerate}
\frac{\omega L^2}{r_h} \approx \frac{n\pi}{\log\left(\frac{L}{\alpha \ell_{p}}\right)},
\eeq
The exact degeneracy assumption, together with a cut-off in $J$, reproduces the correct thermodynamics as well the exterior correlators as already mentioned.  

\subsection{A Bulk Planckian Cut-off}\label{bulkplankian}

\begin{figure}[h]
       \centering
       \includegraphics[width=0.6\linewidth]{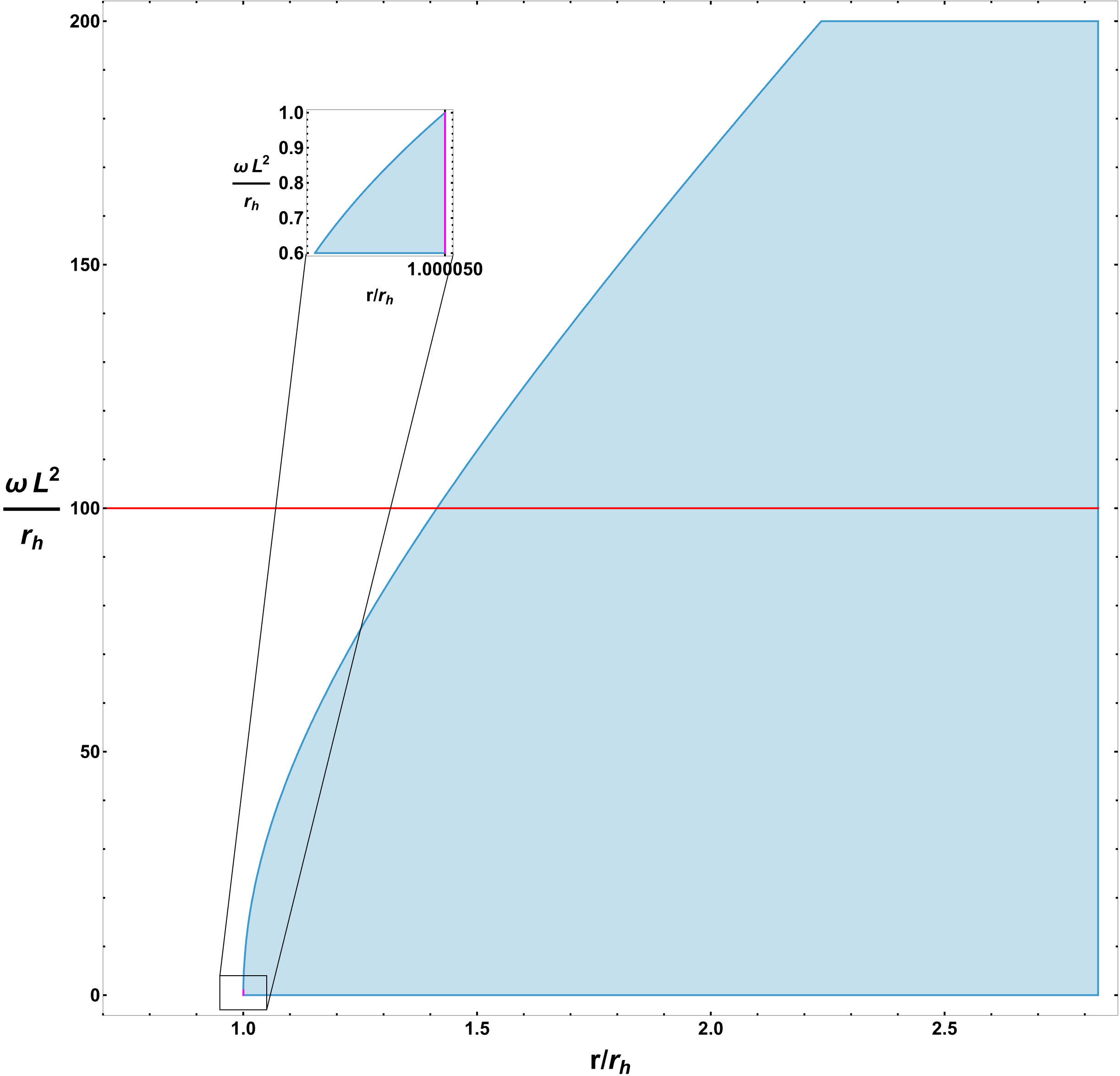}
       \caption{The shaded region satisfies \eqref{omegarplot}. The zoomed-in region is what contributes dominantly to the thermodynamics (as we argue in the text, this corresponds to $\frac{\omega L^2}{r_h} \lesssim \mathcal{O}(1)$): the pink line represents the radial region bounded by \eqref{horN} for $\alpha=1$, and we work with $L/\ell_P=100$. Part of the goal of this plot is also to illustrate that the bulk UV cut off is effective only close to the horizon.}
       \label{omegaspecanalys}
\end{figure}

Imposing that the local bulk energy scale does not exceed the bulk UV cut-off, we consider the bound
\beq \label{omegarplot}
\frac{\omega L^2}{r_h} \leq M L 
\sqrt{\left(\frac{r}{r_h}\right)^2 - 1}.
\eeq
This constraint is illustrated in Fig.~\ref{omegaspecanalys}. From this, we extract the effective radial cutoff scale $r_{\epsilon}$ corresponding to the modes that contribute the most to the thermodynamics. In this subsection we discuss some of the associated physics, before turning in the next section to our main goal: the calculation of the exact partition function.

The second-quantized partition function of the fields in thermal equilibrium is~\cite{Burman1, Pradipta1}
\beq \label{secondpartition}
\log Z = -\sum_\omega \log\!\left(1 - e^{-\beta_H \omega}\right),
\eeq
where the ensemble is defined at the Hawking temperature 
\beq
\beta_H = \frac{2\pi L^2}{r_h},
\eeq
of the BTZ black hole. As $\omega$ becomes densely packed in $J$, we pass to the continuum limit (for $n=1$):
\beq
\log Z
=
- \int d\omega \, \rho(\omega)\,
\log\!\left(1 - e^{-\beta_H \omega}\right),
\eeq
where
\beq
\rho(\omega) = \frac{\partial J}{\partial \omega}.
\eeq
The integrand
\beq
- \rho(\omega)\,\log\!\left(1 - e^{-\beta_H \omega}\right)
\eeq
therefore represents the spectral contribution density to $\log Z$ at frequency $\omega$, but along $J$.

\begin{figure}[h]
       \centering
       \includegraphics[width=0.65\linewidth]{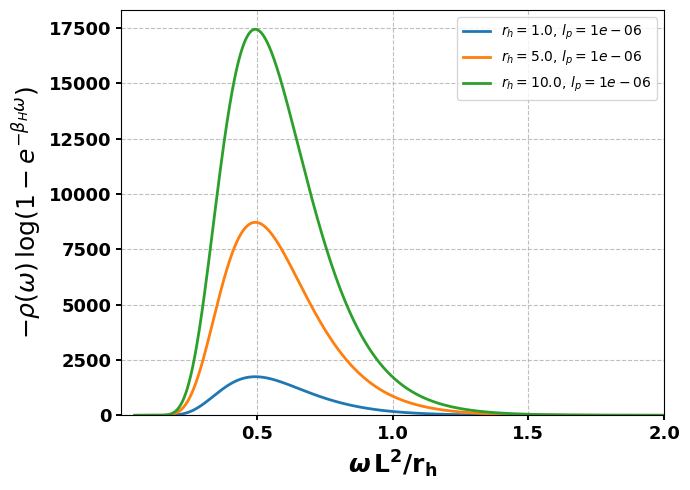}
       \caption{Spectral contribution density $-\rho(\omega)\log(1-e^{-\beta_H\omega})$ for $n=1$ and $\alpha=1$.}
       \label{contributionOfOmega}
\end{figure}

The approximate spectrum $\omega(J)$ obtained in Eq.~\eqref{BTZALLS}, with the digammas approximated by logarithms in $J$, can be inverted to determine $J(\omega)$ and hence the density of states $\rho(\omega)$. For $n=1$, this yields
\beq
\rho(\omega)\,\log\!\left(1 - e^{-\beta_H \omega}\right) =
\frac{e^{-\pi/\widetilde\omega} \, r_h}{\alpha\ell_p}
\left(\frac{\pi}{\widetilde \omega^3} - \frac{1}{\widetilde \omega^2} \right) \log\!\left(1 - e^{-2 \pi \widetilde \omega} \right) \ , \quad {\rm where} \ \widetilde \omega = \frac{\omega L^2}{r_h} 
\eeq
The corresponding profile is displayed in Fig.~\ref{contributionOfOmega}. The distribution exhibits a pronounced maximum near
\beq \nonumber
\frac{\omega L^2}{r_h} \simeq 0.5,
\eeq
independent of $r_h$, $L$, and $\ell_p$. 

Although a closed-form expression for the exact spectrum is not available, comparison between the {\em ALLS}, for the holographically anchored case, and the numerically determined exact spectrum (Fig.~\ref{HSHspectrum}) shows good agreement up to 
\[
\frac{\omega L^2}{r_h} \sim 0.8,
\]
indicating that the dominant thermodynamic contribution is robust against higher-order corrections.
We therefore conclude that the dominant thermodynamic contribution arises from frequencies satisfying
\beq
\frac{\omega L^2}{r_h} \sim \mathcal{O}(1).
\eeq
Using \eqref{boundarydef}, this translates to the radial cut-off
\beq \label{horN}
\frac{r_\epsilon}{r_h} 
\sim 
\sqrt{1+\frac{\ell_p^2}{L^2}}
\eeq
where we have set $\alpha = 1$.

\section{Exact Thermodynamics} \label{sec:exactthermo}

In this section we compute the exact thermodynamics of the normal modes. In the analytic treatment in the hardwall case, the spectrum is often approximated to be more degenerate than it is \cite{tHooft, Burman1, Burman2}. This makes computations of the partition function, thermodynamics, and correlators analytically tractable. But an exact degeneracy in $J$ leads to divergences because of the infinite range of $J$. These divergences are then regulated by introducing a cutoff $J_{\text{cut}}$. Notably, such a $J_{\rm cut}$ can be obtained by the demand that only the classically\footnote{The terminology “classically allowed” here is slightly misleading, even though we use it because it is convenient. What is actually used (see Appendix~\ref{sec:Jcut}) is the fully quantum Sturm-Liouville fact that, after passing to the standard Schr\"odinger form $-u''+V_J\,u=\omega^2 u$ (with $u(r_\epsilon)=0$ at the brick wall and normalizability at infinity), there is \emph{no} nontrivial solution if $\omega^2<V_J(r)$ for \emph{all} $r\ge r_\epsilon$. This follows straightforwardly if we multiply by $u^*$ and integrate by parts to get $\int (|u'|^2+(V_J-\omega^2)|u|^2)=0$. Of course a legitimate mode in a general quantum system may have regions with $\omega^2<V_J$ (tunneling): what is excluded is being “under the potential” \emph{everywhere}. This is in fact what 't Hooft's semi-classicality demand \cite{tHooft} also accomplishes: its real virtue is that the semi-classical spectrum is more degenerate than the exact spectrum, so that it goes under the potential for large enough $J$.} allowed modes are to be retained. With such a choice, one reproduces the Hawking temperature (if the ensemble is specified by the black hole mass), the expected Bekenstein-Hawking entropy, as well as exterior Hartle-Hawking correlators (together with exponentially suppressed corrections) \cite{Burman1,Burman2}.

The natural question is: what happens if we abandon this approximation and instead use the exact spectrum obtained from solving the wave equation, with no degenerate approximation and no imposed cutoff? Analytically this is difficult, since even the analytic low-lying spectrum (reliable up to $\omega \sim r_h/L^2$) \eqref{BTZhardwall} does not lead to a simple closed-form expression for the partition function. We therefore proceed numerically and compute the entropy $S$ and average energy $U$ in the canonical ensemble, fixing the temperature to be the Hawking temperature of the BTZ black hole.

To quantify the results, let us first note the geometric entropy and mass:
\beq
S_{BTZ} = \frac{\pi r_h}{2 G_N},
\qquad
M_{BTZ} = \frac{r_h^2}{8 G_N L^2}.
\eeq
These are to be compared with the entropy $S$ and average energy $U$ obtained from the numerical computation of the exact spectrum. The energy is computed from the partition function as
\beq
U = -\frac{\partial}{\partial \beta} \log Z
  = \sum_{n,J} \frac{\omega_{n,J}}{e^{\beta \omega_{n,J}} - 1},
\eeq
and the entropy is given by
\beq
S = \beta U + \log Z
  = \sum_{n,J} \left[
      \frac{\beta \omega_{n,J}}{e^{\beta \omega_{n,J}} - 1}
      - \log\!\left(1 - e^{-\beta \omega_{n,J}}\right)
    \right].
\eeq
Here $\omega_{n,J}$ denotes the spectrum obtained numerically from the wave equation \eqref{BTZexactphase} as we vary ${\ell}_p$. Note that the discussion so far applies equally well to the horizon-anchored or boundary-anchored definitions of the normal modes.

Before presenting the results, let us summarize the parameters entering the computation and the numerical strategy. The ensemble is defined by fixing
\beq
\beta = \beta_H = \frac{2\pi L^2}{r_h},
\eeq
and the spectrum $\omega_{n,J}$ is obtained numerically from Eq.~\eqref{BTZexactphase}. We will start our discussion with the exact spectrum for the holographically anchored case. Altogether, the independent parameters are $L, r_h, J, \Delta, \alpha, \ell_p,$ and $\omega$. We will set $\Delta=2$ which means that the scalar is massless -- even though we have not attempted an exhaustive study, we have noticed that changing the mass of the scalar does not change the physics as long as it is hierarchically smaller than the UV scales. In all numerical computations we set $L=1$, this is just a choice of unit. The parameters $\alpha$ and $\ell_p$ always appear in the combination $\alpha \ell_p (= \widetilde \ell)$ in the statistical mechanics of normal modes: changing $\alpha$ effectively corresponds to moving the wall to trans-Planckian or sub-Planckian scales. The angular momentum $J$ takes integer values.

A crucial structural feature of the exact spectrum is that it rises with $J$, even though the dependence is weak. As $J$ increases, the frequencies grow and higher modes become energetically expensive. In the canonical ensemble these modes are exponentially suppressed. Consequently, the partition function, entropy, and energy saturate naturally as one sums over increasing $(n,J)$. Unlike the degenerate approximation, no explicit $J_{\text{cut}}$ is required to render the sums finite.

We first fix $r_h=1$ and $\alpha=1$, and vary $\ell_p$. The idea in setting $\alpha=1$ is that we are starting with the maximum allowed bulk UV cut-off, namely the cut-off being equal to the Planck scale\footnote{Our eventual point will be that this is not sufficient to get a satisfactory match with black hole thermodynamics: we will need the UV cut-off to be trans-Planckian by a small hierarchy.}. For each choice of $\ell_p$, we compute the spectrum and evaluate the entropy and energy. Since we are working in $2+1$ dimensions, we identify $G_N=\ell_p$ in the expressions for $S_{BTZ}$ and $M_{BTZ}$.

\begin{figure}
\centering
\begin{subfigure}{.5\textwidth}
  \centering
  \includegraphics[width=0.9\linewidth]{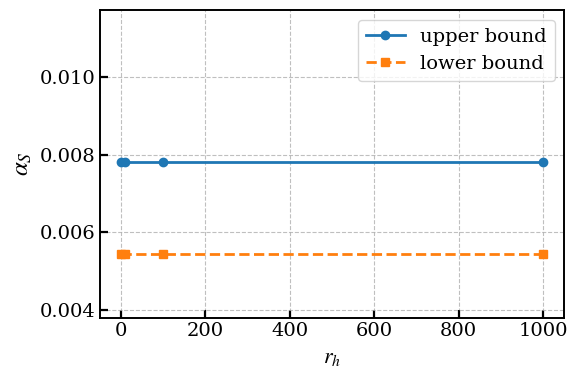}
  \caption{}
  \label{}
\end{subfigure}%
\begin{subfigure}{.5\textwidth}
  \centering
  \includegraphics[width=0.9\linewidth]{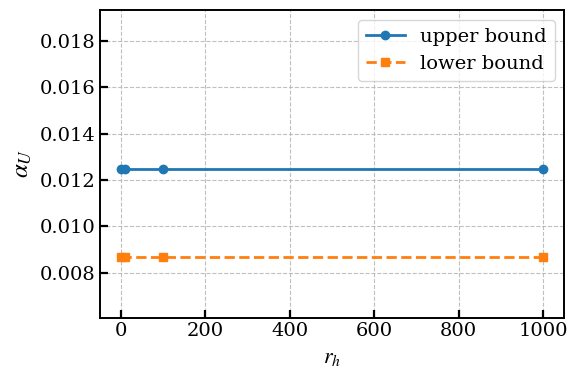}
  \caption{}
  \label{}
\end{subfigure}
\caption{Plot of $\alpha$ v/s $r_h$ for (a) entropy and (b) energy with exact spectrum, for $\ell_p = 10^{-16}$. The $\alpha$'s are defined in \eqref{eq:alpha_S_alpha_U_def}.}
\label{SUratioexactwithrh}
\end{figure}

Once the saturation of the mode sum in the partition function is established for a given $\ell_p$, we test area scaling of the entropy by varying $r_h$. We have checked that $S/S_{BTZ}$ and $U/M_{BTZ}$ remain stable as we change $r_h$ for small enough $\ell_p$, see Fig. \ref{SUratioexactwithrh}. We select a value of $\ell_p$ for which the ratio $S/S_{BTZ}$ and $U/M_{BTZ}$ have stabilized. To claim that this set up can successfully reproduce black hole physics, we will need these ratios to stabilize to a number that is order one or higher, for small $\ell_p$. If it is precisely unity, it would mean that the stretched horizon needs to be precisely at a Planck length for it to reproduce black hole thermodynamics. If the ratios are bigger, we can reproduce black hole physics by moving the cut off back from $\ell_p$ to a bigger (i.e., sub-Planckian, say string length) value.

We define the dimensionless ratios
\beq
\alpha_S = \frac{S}{S_{BTZ}},
\qquad
\alpha_U = \frac{U}{M_{BTZ}}.
\label{eq:alpha_S_alpha_U_def}
\eeq
The ratios $\alpha_S$ and $\alpha_U$ are the outputs of the numerical calculation. If they are not equal to unity (and they are not, if we start with $\alpha=1$), we can now re-do the exact numerical calculation with $\alpha$ replaced by $\alpha_S$ (or $\alpha_U$). When we do this, we find that the new value of $\alpha_S$ (or $\alpha_U$) that results from the calculation is unity, and the $\alpha_U$ (resp. $\alpha_S$) is $\mathcal{O}(1)$. This is a quantitative way of stating our comments in the previous paragraph. It is important for this discussion to make physical sense, that the values of $\alpha_S$ and $\alpha_U$ that we obtain when we start with $\alpha=1$ are greater than or equal to $\mathcal{O}(1)$: if not, it will mean that we are forced to set the UV cut off to be trans-Planckian to reproduce black hole thermodynamics.

\begin{figure}
\centering
\begin{subfigure}{.5\textwidth}
  \centering
  \includegraphics[width=0.9\linewidth]{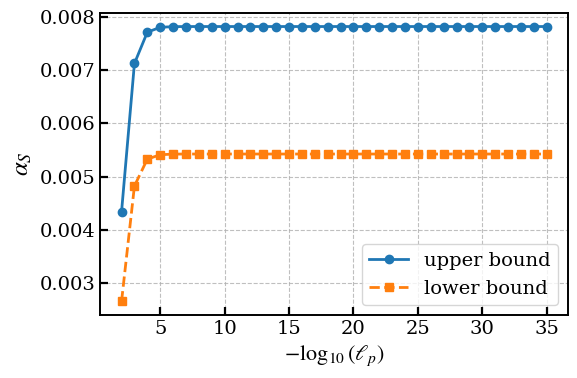}
  \caption{}
  \label{}
\end{subfigure}%
\begin{subfigure}{.5\textwidth}
  \centering
  \includegraphics[width=0.9\linewidth]{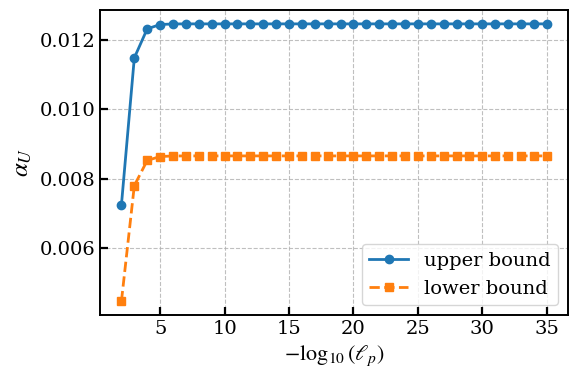}
  \caption{}
  \label{}
\end{subfigure}
\caption{Plot of $\alpha$ v/s $\ell_p$ for (a) entropy and (b) energy with exact spectrum, for $r_h/L = 1$. The $\alpha$'s are defined in \eqref{eq:alpha_S_alpha_U_def}.}
\label{SUratioexact}
\end{figure}

The results of the exact numerical calculation of $\alpha_S$ and $\alpha_U$ are shown in Fig.~\ref{SUratioexact} and Fig. \ref{SUratioexactwithrh}. We find that entropy and energy both scale correctly with the black hole radius (the former, as the area) confirming that the exact spectrum reproduces area scaling. However, a hierarchy appears in the overall coefficients. As clear from the figures, when we start with $\alpha=1$ in the calculation, we obtain
\beq
\alpha_S \sim 10^{-3},
\qquad
\alpha_U \sim 10^{-2}.
\eeq
Thus entropy and energy are suppressed relative to the geometric values, and moreover they are not suppressed by the same factor. This persists even when $\ell_p$ is decreased to smaller values and is insensitive to changes in $r_h/L$. We have checked that a hierarchy of the same order of $\alpha_S$ and $\alpha_U$ is present when working with the hard-cutoff brick wall spectrum used in \cite{tHooft, Burman1, Burman2} as well -- so this result is not an artifact of our holographic-anchoring. We have also checked that similar (in fact, worse) hierarchies exist even for other black holes (e.g., Schwarzschild or Reissner-Nordstrom in 3+1 dimensions).

It is worth noting that although \eqref{BTZexactphase} holds at any radial location, the entropy does not obey area scaling when the bulk radial cutoff is far from the horizon: $\alpha_S$ and $\alpha_U$ \eqref{eq:alpha_S_alpha_U_def} scale linearly with $\ell_p$ in Fig.~\ref{SUratioexactwithrh}. The numerical results saturate at small $\ell_p$ (i.e., when $\ell_p$ is much smaller than all other scales in the problem) because of the near-horizon redshift, which is the underlying mechanism for area scaling. Hence, for sufficiently small $\ell_p$ the scaling of $\alpha_S$ and $\alpha_U$ with $r_h$ can be meaningfully examined.

The conclusion that emerges from our numerical computations with spectra that have nontrivial $J$-dependence is the following. Dropping (or ameliorating) that dependence, and assuming that the spectrum is essentially degenerate, was crucial for the success of the previous calculations in reproducing black hole physics. 

We emphasize that this conclusion is based on direct numerical computations and estimates of the partition function. But while we do not have an analytic proof, we believe the result is reasonable. The area scaling of the entropy is a consequence of the fact that one of the quantum numbers that are available in flat space are unavailable near the horizon: a missing quantum number effectively reduces the effective dimensionality of the spacetime and we get area- instead of volume-scaling. The trouble with a weak $J$-dependence is that this argument is not quite correct. Once the spectrum genuinely increases with $J$—as in the exact spectrum obtained from solving the wave equation—it no longer seems possible to match both entropy and energy to their BTZ values at the same time. It is intuitive from the structure of the spectrum (and the fact that it is not {\em that} degenerate) that the exact calculation should carry some vestige of that broken degeneracy. We believe the hierarchy as well as the mismatch between $\alpha_S$ and $\alpha_U$ are traces of this fact in the statistical mechanics.

If we compare these results with the original WKB analysis of 't Hooft, we find that an exact matching between statistical and geometric quantities does not occur automatically there either (see Appendix \ref{sec:tHooft}). Matching either entropy or energy requires choosing the proper distance cutoff to be a small fraction of the Planck length. In the present language, this is equivalent to taking the parameter $\alpha$ in \eqref{boundarydef} to be of order $10^{-1}-10^{-3}$ depending on the black hole, anchoring choice, etc. We demonstrate this in Appendix \ref{sec:tHooft}. In other words, the hierarchy we observe in doing the computation with the exact numerical spectrum was implicit in the semiclassical treatment as well -- except that the hierarchy was slightly weaker there, because the semi-classical spectrum is more degenerate than the exact spectrum. 

What we are emphasizing in this section is that one cannot wish this away as an artifact of the WKB analysis: it is present even when the computation is performed with the exact normal modes, and in fact the hierarchy gets worse. Interpreted in the hard cutoff picture, the hierarchy indicates the need for placing the wall at a shorter-than-Planckian proper distance from the horizon.

\subsection{Comments on the Numerics}

\begin{figure}[h]
       \centering
       \includegraphics[width=0.65\linewidth]{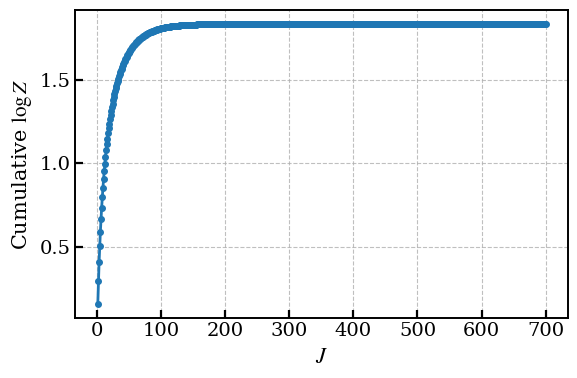}
       \caption{Saturation of the partition function for $\ell_p=10^{-3}, r_h=1, \alpha=1$, with the sum taken over $n=1$ to $5$.}
       \label{partiton_function_saturation}
\end{figure}

In practice, we compute the spectrum numerically by summing over $n=1$ to $5$\footnote{The numerical sum saturates rapidly in $n$, with the dominant contribution coming from $n=1$.}, while taking $J$ in a wide symmetric range. Although the formal limit over $J$ is infinite, a saturation in the numerical sum typically occurs at 
\beq 
J=J_{\rm sat} \lesssim \frac{r_h}{{\alpha\ell}_p} 
\eeq 
Extending the range further does not change the thermodynamic quantities appreciably.
Importantly, the exact thermodynamics saturates naturally without the need to impose any explicit cutoff in the angular momentum. The $J_{\rm sat}$ defined above is of the same order as the value of $J$ at which the {\em ALLS} (\ref{sliverspec}) diverges, namely $J_{\rm max} \equiv \frac{r_h}{\alpha \ell_p}$, making it a natural and convenient reference scale. We have explicitly verified that the thermodynamic quantities saturate well before this value. For instance, in Fig.~\ref{partiton_function_saturation}, we take $J_{\rm max}=1000$ for the chosen parameters, while the numerical sum saturates at much smaller values of $J$.

A technical difficulty in the exact computation is that $J_{\rm sat} \sim \frac{r_h}{\alpha\ell_p}$
becomes large for small $\ell_p$. For values such as $\ell_p \sim 10^{-20}$ or $\ell_p \sim 10^{-30}$, the number of $J$ modes becomes difficult to manage and a brute-force summation over all integer $J$ is computationally unfeasible. Instead, we perform controlled sampling in $J$. The spectrum is computed on a sparse grid of $J$ values, and for intermediate integer values not explicitly sampled, the corresponding $\omega$ is assigned from the nearest available sampled value, either from the previous or the next sampled $J$. This procedure provides upper and lower bounds on the thermodynamic quantities. The difference between these bounds can be systematically reduced by increasing the sampling density.

Appendix~\ref{sec:sliver} describes one implementation of the sampling strategy, in which the sampled $J$ values are determined by dividing the $\omega$-axis into equal segments. Such a procedure is useful both numerically and analytically, especially if the spectrum can be inverted to express $J$ as a function of $\omega$, leading to a simple functional form such as $g(J_{\max},n)$ (defined in Appendix~\ref{sec:sliver}). Thus, whenever we refer to the “exact” computation, we mean that the exact numerical dispersion relation is used, while the summation over $J$ is performed using controlled sampling, yielding well-defined upper and lower bounds on the computed thermodynamic quantities.

In the numerical computation with the exact spectrum, we use the following sampling scheme:
\beq
\mathcal{J}
=
\bigcup_{k=0}^{k_{\max}}
\left\{
J = b^k m
\;\middle|\;
m=1,2,\dots,b-1,
\; b^k m \le J_{\rm max}
\right\}.
\eeq
Fig. \ref{SUratioexact} and \ref{SUratioexactwithrh} correspond to the choice $b=10$. Other values of $b$ can also be used. Choosing a denser sampling systematically reduces the gap between the upper and lower bounds. For certain intermediate values of $b$, mild fluctuations of the bound curves may be observed; however, they remain confined within the $b=10$ bound band. These variations originate from the coarse sub-sampling of $J$ leading to minor sampling artifacts in the extracted bounds. In the limit of large $b$ (corresponding to very dense sampling), the two bounds approach each other. Performing such dense sampling is computationally expensive -- but it can be verified for $\ell_p$ values that are not too small (say, $\ell_p$ up to $10^{-8}L$), that the two bounds tend towards each others and merge as the sampling becomes denser.

To summarize: we can compute the exact partition function only up to some not too small values of $\ell_p$. But for those values, we can check by dense sampling that the upper and lower bounds also converge to that value from above and below. For smaller values of $\ell_p$ where we cannot do the exact calculation because of the proliferation of points, we can {\em still} check that the upper and lower bounds stabilize and become constants once the $\ell_p$ is small enough. This is one of the messages of Figure \ref{SUratioexact}. With denser sampling, we can also see that the band between the upper and lower bounds shrinks. What we cannot do, is to show that the band shrinks down to (essentially) zero size for arbitrarily small $\ell_p$ because this requires a sampling too dense to be numerically feasible. But this is sufficient for our purposes because the width of the band is small compared to the hierarchy that we wish to establish.

Finally, the numerical observations also show that the {\em ALLS} spectrum provides a reliable approximation for thermodynamic purposes. Since the dominant contribution to the partition function arises from the same region of the spectrum, both horizon-anchored and boundary-anchored walls yield qualitatively similar thermodynamic behavior, including the small but persistent hierarchy in the overall coefficients. In other words, the {\em ALLS} results are perfectly consistent with the ``bound"-based exact calculations. 

In summary, the exact-spectrum computation shows that the thermodynamic sums saturate naturally, the entropy scales with the horizon area, and a persistent small coefficient relates microscopic counting to geometric quantities.

\section{Bulk Sources as Microstates}

What are the possible ways to mitigate the mini-hierarchy that we have seen in this paper? One possibility is to add more active species (fields) at the cut-off scale. This is certainly a possibility in string theory, and a few hundred species (including spins) is not a wild possibility in typical BSM models. But this still falls short of a satisfactory resolution because of the $\mathcal{O}(1)$ mismatch, when we try to fix not just the entropy, but also the mass (energy). The holographically anchored stretched horizon leads to marginally better conclusions on both the hierarchy and the mismatch -- but not enough to solve the problem entirely.

We will view these observations as indicating that the $J$-degenerate spectrum considered in \cite{Burman1, Burman2} (and in related papers) is, as anticipated there, a ``toy model". The success of those calculations relied on the toy model spectrum being discrete, and it being able to reproduce the Hawking temperature and the Bekenstein-Hawking entropy precisely, once the ensemble was fixed via the mass/energy of the black hole\footnote{Said differently: the approximate $J$-degeneracy of the brick wall spectrum lead to the qualitative area law success in \cite{tHooft}, which allowed us to guess a toy spectrum which was {\em quantitatively} successful \cite{Burman1, Burman2}. The results of this paper show that the raw brick wall spectrum, without this by-hand ``improvement",  struggles to reproduce the quantitative features of black holes.}. We expect that {\em any} spectrum that has these features would be able to reproduce the correct Hartle-Hawking correlators in a suitable large-$N$ limit. But to build more realistic models of microstates, we will need to incorporate more realistic UV inputs. What would such a construction entail? 

In AdS$_3$/CFT$_2$, we can shed some light on this question by taking inspiration from some of the recent results in \cite{CKPSP} (see also \cite{Datar}). The idea here is that a bulk interpretation can be given to the following well-known CFT fact: the vacuum character in the dual channel can be written as a sum over characters with the modular S-kernel as the weight \cite{Ponsot}. In the bulk, this provides us an understanding of the BTZ character -- the latter can be written as the sum over characters of super-threshold primaries, with the S-kernel serving as the density of primaries. In the Chern-Simons language, the microstates are sourced by Wilson lines with hyperbolic holonomy, placed at the origin of the spatial disc along the Euclidean time direction of the solid cylinder. In the metric language, individual states of this kind correspond to BTZ geometries with singular horizons (because the periodicity in the thermal direction can be arbitrary)\footnote{The non-vanishing hyperbolic holonomy of the Wilson line source translates to the fact that the black hole microstate has a finite ``horizon" size, even though in the Chern-Simons language the Wilson line is sitting at the origin.}. Despite this, the $S$-kernel {\em sum} over primaries is the partition function of a {\em smooth} BTZ handlebody. 

In this picture, each individual microstate has a singular horizon\footnote{By (absence of) analytic continuation, this is true in Lorentzian signature as well.}, but outside the horizon the metric description is locally AdS$_3$: indeed it is smooth BTZ everywhere outside the horizon. But note that unlike the usual BTZ black hole which is supposed to be understood as an ensemble of microstates with some fuzziness in the parameters (say $r_+$ and $r_-$), here we have a unique state with the parameters precisely fixed: the $r_+$ and $r_-$ translates to the $h$ and $\bar h$ that specify the primaries (microstates) in the dual CFT. This picture of microstates has the virtue that each microstate shares the underlying $U(1)$ isometry of the BTZ black hole\footnote{In higher dimensions, this is consistent with the expectation that microstates share the isometries of the black hole \cite{Sen}.}. 

These observations suggest the following  microscopic model for a light scalar interacting with the black hole. The natural language to work with is the Chern-Simons formulation, with a heavy Wilson line sourcing the microstate. The scalar field probing the microstate via a 2-point correlator in this heavy background is to be modeled using the interaction of the core Wilson line with the scalar Wilson line anchored to the boundary. This picture allows a direct connection to the CFT via the Virasoro block representation of the HLLH correlators \cite{Kaplan}. It is natural to suspect that an S-kernel like sum over the heavy primaries will be equal (in the dual channel) to the light scalar correlator in the smooth BTZ handlebody. We hope to report on this picture in upcoming work \cite{VishalVaibhav}. Considering the fact that the S-kernel formula for characters holds at finite-$c$, there is a chance that this approach may reveal more than what is known from semi-classical limits of Virasoro blocks: the latter is simply the statement that HLLH correlators on {\em individual} heavy microstates reduce to thermal LL correlators in the large-$c$ limit. (Note that Cardy formula holds even at finite-$c$ for high temperatures.)

In the full UV-complete picture, the discreteness of the microstates is a statement about the holographic CFT: a discrete spectrum needs to be compatible with modular bootstrap like arguments \cite{CKPSP}. The S-kernel itself is continuous and captures the modular transformation properties of the characters, but modular invariance of the CFT partition function is realized via its discrete spectrum of primaries \cite{CKPSP}. The {\em averaged} density of states on the other hand, is captured by the continuum S-kernel. From the perspective of the $S$-kernel or bulk EFT, the discreteness is a highly non-trivial result of UV consistency \cite{CKPSP}. This ``magical" origin of the CFT discrete spectrum is what is being imperfectly simulated by crude brick wall like toy models through Dirichlet (or similar) boundary conditions at a finite distance outside the horizon. 

\subsection{Spectral Correlations vs. Spectral Density}\label{sec:correlations}

Given the conclusions of this paper, an important question to address is the following. It has been observed in \cite{Adepu, Sumit, ramp} that the brick wall spectrum -- in particular the slow logarithmic growth of $\omega$ with $J$ -- reproduces various features expected of random matrices and quantum chaos: the spectral form factor (SFF) exhibits a dip-ramp-plateau structure with a linear ramp of slope $\sim 1$, and level repulsion emerges once a small noise correction (motivated by fuzzball profile functions) is added to the spectrum. Furthermore, the simplest spectrum exhibiting a linear ramp is $E_n \sim \log n$ \cite{ramp}, and the SFF of this spectrum is the mod square of the Riemann zeta function. This observation was recently exploited in \cite{Pallab} to analytically establish a linear ramp with slope precisely equal to $1$, and to demonstrate that the Thouless time is $\mathcal{O}(1)$ -- a property expected of strongly chaotic systems such as black holes. If the brick wall spectrum is ``merely" a toy model, as the little hierarchy of this paper suggests, why does it exhibit such non-trivial signatures of quantum chaos?

We believe there is a clean resolution, which rests on a distinction between {\em spectral density} and {\em spectral correlations}. The little hierarchy identified in this paper is a statement about the spectral density: the overall {\em coefficient} of the area law. The non-trivial $J$-dependence means there are not quite enough states to match the Bekenstein-Hawking entropy, unless the brick wall is trans-Planckian. This is a problem about getting the right {\em number} of microstates. In contrast, the random matrix diagnostics -- the linear ramp, level repulsion, the $\mathcal{O}(1)$ Thouless time -- are properties of spectral {\em correlations} rather than the overall density of states. Let us also note a very heuristic analogue for this in random matrix theory: the {\em unfolded} spectrum (after removing the smooth density of states) has universal correlations determined by the symmetry class, while the density of states itself is system-specific and non-universal. The brick wall appears to be getting the former right while missing the latter.

The reason why the brick wall gets the spectral correlations right is that the weakly logarithmic $J$-dependence is a robust consequence of the near-horizon geometry\footnote{On the other hand, why logarithmic spectra seem to know about RMT spectral correlations is something that does not seem to be sufficiently explored in the current literature.}. The underlying mechanism is a scale-invariance property of the near-horizon region. Let us be precise about its nature, and how it differs from other instances of scale-invariance, like in inflation. In de Sitter space, the dilation $\eta \to \lambda \eta, \, \mathbf{x} \to \lambda \mathbf{x}$ is an exact isometry of the metric $ds^2 = (H^2 \eta^2)^{-1}(-d\eta^2 + d\mathbf{x}^2)$, and the observable -- the perturbation amplitude -- is a {\em local} quantity evaluated at each mode's horizon crossing ($k|\eta| \sim 1$). Because the geometry at horizon crossing is the same for every $k$ (exact isometry), the output is a scale-invariant power law $P(k) \sim k^{n_s - 1}$: power laws are the natural eigenfunctions of the dilation operator $k \partial_k$. 

In the black hole context on the other hand, the near-horizon metric $f(r) \approx 2\kappa(r - r_h)$ gives rise to a tortoise coordinate integrand $dr/f(r) \approx dr/[2\kappa(r-r_h)]$ that is scale-invariant under $r - r_h \to \lambda(r - r_h)$. However, this is {\em not} an isometry of the full metric. The angular part $r^2 d\Omega^2$ breaks it. But it is a self-similarity of the radial throat encoded in the simple pole of $1/f$ at the horizon. Moreover, the spectrum $\omega_{n,J}$ is not a local amplitude but a {\em global} quantity: the total accumulated WKB phase $\int dr/f(r)$ from the wall to the turning point. Integrating a scale-free integrand over a range of scales from a UV cut-off $\epsilon$ to an IR scale gives a logarithm, $\int dx/x = \log x$. This can be seen explicitly in the 't Hooftian semi-classical spectrum \eqref{thooftspectrum} and in the ALLS \eqref{BTZALLS}, \eqref{BTZhardwall}. The spectrum takes the schematic form
\beq
\omega \sim \frac{n\pi}{\log\left({L}/{\alpha \ell_p}\right) - \log\left({JL}/{r_h}\right) + \ldots} = \frac{n\pi}{\log\left(J_{\rm max}/J\right) + \ldots}
\eeq
where $J_{\rm max} = r_h/(\alpha \ell_p)$, and the slow logarithmic growth in $J$ -- responsible for the linear ramp, level repulsion (once there is slight noise), the zeta function connection, and the $\mathcal{O}(1)$ Thouless time -- arises from the self-similarity of the near-horizon throat. 

This picture leads to a suggestive expectation for the more realistic microstate models alluded to earlier in this section, where the dominant degeneracy arises from a radial quantum number (as suggested by the $U(1)$ isometry of the microstates in \cite{CKPSP}). In such a model, the intrinsic horizon degrees of freedom would fix the spectral {\em density} -- resolving the little hierarchy. The $J$-dependence of the spectrum, being a universal consequence of the near-horizon geometry, would still be present. In the brick wall toy model, it is this $J$-dependence that provides the spectral correlations responsible for the random matrix diagnostics \cite{Adepu, Sumit, ramp, Pallab}. Whether it continues to dominate the spectral correlations once the radial degrees of freedom are properly included is an open question: the discrete ``radial" spectrum could itself introduce non-trivial level statistics. 

\subsubsection{A Hagedorn Connection}

The spectral density of the brick wall model $\rho(\omega) = dJ/d\omega$, follows from inverting the full\footnote{The discussion here is heuristic, and we are only interested in the low-lying part of the spectrum where analytic approximations are available. The form here captures the crucial features of the {\em ALLS}, see the discussion in \cite{Pradipta1} for more details.} spectrum $\omega(J) \sim 1/\log(J_{\rm max}/J)$. This gives $J(\omega) = J_{\rm max}\, e^{-c/\omega}$ and hence $\rho(\omega) \sim (J_{\rm max}/\omega^2)\, e^{-c/\omega}$, which is exponentially {\em suppressed} at small $\omega$ and peaks at $\widetilde \omega \sim \mathcal{O}(1)$, as computed explicitly in Section~\ref{bulkplankian}. The area law for entropy in the brick wall case arises {\em not} from a Hagedorn-like growth of the density, 
but from the near-horizon redshift, with the coefficient being fixed by
$J_{\rm sat} \sim J_{\rm max} \sim r_h/(\alpha\ell_p)$. 
The {\em precise} coefficient -- which is what the little hierarchy is about -- presumably depends on the details of the  cut-off.\footnote{We remind the reader that in the exactly degenerate approximation, we can get a precise match with the Bekenstein-Hawking coefficient by choosing $J_{\rm cut}$ \cite{Burman1}.} The Hagedorn growth of states is usually associated to the perturbative string limit, so these observations about the brick wall spectrum are not in direct tension with the lore about black hole-string transition \cite{SusskindBHstring, Polchinski}.

There is, however, a separate but related connection to Hagedorn physics, which operates at the level of spectral correlations rather than the thermodynamic density. The {\em correction} to the degenerate spectrum grows as $\delta\omega \propto \log J$ for $J \ll J_{\rm max}$, so the SFF-relevant part of the spectrum takes the form $E_n \sim \log n$. The partition function of this toy spectrum is $\sum_{n=1}^{N} n^{-s} \to \zeta(s)$, and the $s=1$ pole of the Riemann zeta function can be interpreted \cite{Pallab} as a Hagedorn transition of the toy model (since the density of states of $E_n = \log n$ is $\rho(E) = e^E$, which is genuinely Hagedorn). This is what connects the near-horizon geometry to the analytic zeta function ramp and the $\mathcal{O}(1)$ Thouless time established in \cite{Pallab}. In other words, the Hagedorn-like physics is visible in the spectral correlations of the brick wall (which see the $\log n$ behavior and hence the zeta function), but not in the thermodynamic density of states (which sees the full $1/\log(J_{\rm max}/J)$ and is controlled by $J_{\rm max}$). 

This reinforces the distinction drawn in the previous subsection: spectral correlations are related to near-horizon universality, while the spectral density requires UV input.

\subsection{Lessons for Fuzzballs?}

The fuzzball program \cite{Mathur, BenaWarner} posits that the $e^S$ microstates of a black hole are realized as smooth horizonless ``geometries" in string theory. Its most complete success remains the 2-charge D1-D5 system, where geometric quantization of the full supergravity moduli space precisely reproduces the entropy \cite{KST, KrishnanRaju, Rychkov}. However, for black holes with finite classical horizon area, such as D1-D5-P, the known supergravity microstate geometries account for only a parametrically subleading fraction of the Bekenstein-Hawking entropy\footnote{The analysis of \cite{BenaWarnerEntropy} argues that shape modes of the superstratum in the non-compact directions account for at least $\sim 1/\sqrt{6}$ of the D1-D5-P entropy.}. The recent monotone/fortuitous classification of BPS cohomologies \cite{ChangLin, ChangLinZhang, CKSLY, ChangSiaYang, HughesShigemori} provides a structural explanation: the smooth horizonless SUGRA solutions that the fuzzball program constructs correspond to monotone states, which are vastly outnumbered by fortuitous states -- finite-$N$ entities with no (smooth) supergravity description.

The results of this paper and \cite{CKPSP} fit naturally into this picture:
\begin{itemize}
\item In both the fuzzball program and the brick wall, the ``easy'' degrees of freedom -- smooth supergravity solutions (monotone states) and probe scalar modes, respectively -- capture only a subleading fraction of the entropy.
\item The missing degrees of freedom are intrinsically quantum: fortuitous states in the fuzzball context, and the finite-$c$ discrete spectrum selected by modular invariance as in \cite{CKPSP}. In neither case do the individual microstates admit smooth classical descriptions.
\item The microstates in \cite{CKPSP} -- BTZ geometries with singular horizons\footnote{Let us note a slight tension in adjectives. The title of \cite{CKPSP} uses the word ``horizonless'' to describe these singular geometries -- the bulk microstates described in \cite{CKPSP} are manifestly unsmooth. But the fuzzball literature often uses ``horizonless'' to indicate {\em smooth} geometries without a horizon.} whose sum reproduces the smooth BTZ partition function -- provide a concrete realization of singular microstates, aligning with the expectation that typical microstates need not be smooth geometries.
\end{itemize}

\section*{Acknowledgments}
We thank Vaibhav Burman and Suman Das for discussions. V.G. is supported by the Council of Scientific \& Industrial Research (CSIR) Fellowship No. 09/0079(22163)/2025-EMR-I.

\appendix 

\section{The Origin of $J_{\rm cut}$} \label{sec:Jcut}

In this appendix, we present a kinematical argument for the emergence of $J_{\rm cut}$ in the degenerate spectrum. This can be viewed as a small improvement on \cite{Burman1, Burman2} where the cut-off was obtained by simply demanding that the {\em coefficient} of the Bekenstein-Hawking entropy came out right. (Note that the area-scaling itself is a dynamical consequence of the spectrum.)

\begin{figure}[h]
    \centering

    \begin{subfigure}[b]{0.45\textwidth}
        \centering
        \includegraphics[width=\textwidth]{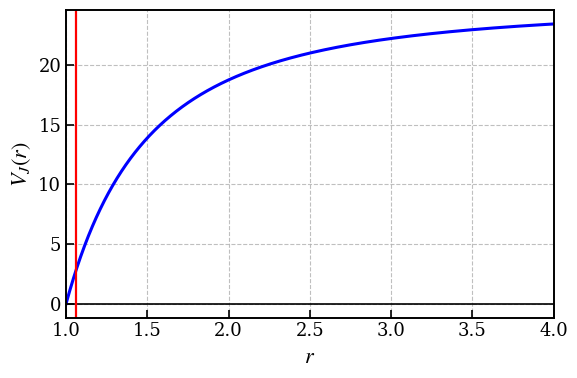}
        \caption{Potential with a brick wall ($J=5$)}
        \label{plot:j=5 stretched horizon}
    \end{subfigure}
    \hfill
    \begin{subfigure}[b]{0.47\textwidth}
        \centering
        \includegraphics[width=\textwidth]{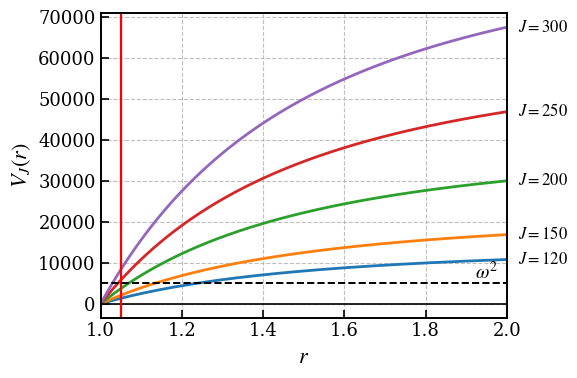}
        \caption{$J_{\rm cut}$}
        \label{plot:l cut}
    \end{subfigure}

    \caption{$J_{\rm cut}$ and Stretched Horizon}
\end{figure}

With the brick wall\footnote{In this discussion, we stick to the horizon-anchored brick wall, which was the case considered in \cite{Burman1, Burman2}.} at a small coordinate distance $\epsilon$ outside the event horizon (shown by the red line in Fig.~\ref{plot:j=5 stretched horizon}), the scalar field is effectively confined in a finite region: one boundary is the stretched horizon and the other boundary is the rising potential at large $r$. Therefore, the field behaves like a particle in a box, and the spectrum becomes discrete.

Now consider the degenerate spectrum approximation \cite{Burman1,Burman2}.  
For a fixed radial quantum number $n$, the energy $\omega$ is taken to be independent of the angular momentum $J$. However, $J$ is a parameter in the potential. The key observation is that as the $J$ keeps increasing, the normal mode energy goes {\em under} the potential. To understand this, draw a horizontal line corresponding to a fixed energy $\omega^2$, see Fig.~\ref{plot:l cut}.  
For small $J$ (for example $J=100$ or $200$ in the figure), part of the potential lies below this energy line, so such modes can exist. But as $J$ increases, the entire potential shifts upward.  At a critical value $J=J_{\text{cut}}$ satisfying
$V_J(r_h+\epsilon)=\omega^2$,
the potential at the stretched horizon itself becomes larger than the $\omega$. For $J>J_{\text{cut}}$ the potential is everywhere above the energy, so this serves as a natural cut-off for $J$. Since the potential monotonically increases with $J$, it never comes back below the energy again. Therefore, only a finite number of angular momentum states contribute.

Let us make this discussion more explicit. For a massless probe near the horizon, $J_{\rm cut}$ can be determined from $\omega^2 =V_{J_{\rm cut}}(r_h+\epsilon)$ 
\beq \label{Jcut}
\omega^2 \simeq \left(\frac{2r_h\epsilon}{L^2}\right)\frac{J^2_{\rm cut}}{r_h^2} \ , \quad \implies J_{\rm cut} \simeq\omega L \sqrt{\frac{r_h}{2\epsilon}},
\eeq
after plugging in the near horizon approximation $r \approx r_h, \  r^2 - r_h^2 \approx 2 r_h \epsilon$, in the potential presented in \cite{Sumit}\footnote{This can be read off from \eqref{radialdiff}, after converting it to the Schrodinger form.}
\beq
V_{J}(r)=\left( \frac{r^2 - r_h^2}{L^2} \right) \left(\frac{J^2}{r^2} + \frac1{L^2} \left(\frac{3}{4} + \frac{r_h^2}{4 r^2} \right) \right). 
\eeq

\noindent
Plugging in the degenerate spectrum \cite{Burman1,Burman2}
\begin{equation}
\omega=\frac{2\pi r_h}{L^2\log \left(\frac{r_h}{2\epsilon}\right)},
\end{equation}
in \eqref{Jcut}, we get: 
\begin{equation}
J_{\rm cut} \approx \frac{2\pi r_h }{L\log\!\left(\frac{r_h}{2\epsilon}\right)}
\sqrt{\frac{r_h}{2\epsilon}},
\label{J_cut_expr}
\end{equation}
For small $\epsilon$, which corresponds to small $G_N /L$, the partition function can written in the degenerate approximation as
\bea
\begin{aligned}
\log Z_{\mathrm{SH}}
&\approx
-\frac{L^{2}\,\log\!\left(\frac{r_h}{2\epsilon}\right)}{2\pi r_h} 
\int_{0}^{\infty}\sum_{J=-J_{\rm cut}}^{J_{\rm cut}}
d\omega\,\log\!\left(1-e^{-\beta_{\rm SH}\omega}\right) ,
\\[2pt]
&\approx \,L^{2}\log\!\left(\frac{r_h}{2\epsilon}\right)\,
\frac{\pi}{6 r_h \beta_{\rm SH}}\,J_{\rm cut},
\end{aligned}
\eea
Using $J_{\rm cut}$ from \eqref{J_cut_expr}, we obtain
\beq
\log Z_{\mathrm{SH}} \approx \frac{\pi^2 L}{3\beta_{\rm SH}}\sqrt{\frac{r_h}{2\epsilon}}.
\eeq
Fixing the canonical ensemble by $\beta_{SH} = \beta_H = \frac{2\pi L^2}{r_h}$ and using $\sqrt{\frac{r_h}{2\epsilon}}=\frac{L}{G_N}$ gives the entropy and average energy as,
\bea
U &=& -\frac{\partial}{\partial\beta_{\rm SH}}\log Z_{\rm SH} = \frac{\pi^{2}L}{3\,\beta_{\rm SH}^{2}}\sqrt{\frac{r_h}{2\epsilon}} = \frac{r_h^{2}}{12L^{2}G_N}, \\
{\rm and,} \quad S &=& \log Z_{\rm  SH} + \beta_{\rm SH}E = \frac{2\pi^{2}L}{3\,\beta_{\rm SH}}\sqrt{\frac{r_h}{2\epsilon}} = \frac{\pi r_h}{3G_N},
\eea
Comparing these with the BTZ Energy/Mass and Entropy
\bea
M &=& \frac{r_h^{2}}{8G_NL^{2}}, \qquad S_{\rm BTZ} = \frac{\pi r_h}{2G_N}, \\ 
{\rm we \ get,} \quad \frac{U}{M} &=& \frac{2}{3}, \qquad \frac{S}{S_{\rm BTZ}} = \frac{2}{3},
\eea
and hence, the entropy and energy/mass match, up to a genuinely $\mathcal O(1)$ number (which is the same for both). 

In fact, it is possible to fix this factor of $2/3$ as well, by viewing the cut-off as being related to the central charge $\big(c=\frac{3L}{2G_N}\big)$ and not directly to the Planck length. To see this, simply fix the brickwall at a geodesic distance \cite{Burman1} equal to $\frac{L}{c}$ instead of $\ell_p \equiv G_N$ from the horizon, i.e. $L\sqrt{\frac{2\epsilon}{r_h}} = \frac{L}{c}$. With that choice, we get a perfect match to the BTZ entropy and mass by repeating the previous argument.

\section{Bounding the Partition Function from Above and Below}\label{sec:sliver}

A significant challenge in the numerical evaluation of the partition function and entropy is the infinite range of the quantum numbers $n$ and $J$. While the sum over $n$ typically saturates rapidly—with the $n=1$ state providing the dominant contribution—the sum over $J$ only saturates at 
\[
J_{\text{max}} = \frac{r_h}{\alpha \ell_p}.
\]
For small values of the Planck length $\ell_p$, $J_{\text{max}}$ becomes extremely large, rendering a direct summation computationally difficult. To address this, we employ a sampling strategy: the frequencies $\omega_{n,J}$ are computed only for a discrete subset of $J$ values, and the contribution of intermediate (unsampled) values is approximated using the nearest sampled frequency. Since contribution to thermodynamic quantities get energy suppression, replacing a missing $\omega_{n,J}$ by a smaller sampled value produces a rigorous upper bound, while using a larger sampled value yields a lower bound.

A generic sampling in $J$ for a fixed $n$ typically leads to non-uniform spacing in the resulting $\omega$'s. This becomes inefficient if the higher $n$ modes also contribute, as the same procedure must then be repeated independently for each $n$. A more efficient approach is to sample $J$ so that the resulting frequencies are equally spaced in units of a fundamental frequency
\beq
\omega_0 \equiv \omega_{1,1}.
\eeq
If the spectrum is linear in $n$, so that $\omega_{n,1}=n\omega_0$ then this choice aligns all $n$-sectors on the same frequency grid. Contributions from different $n$ can then be grouped within common frequency intervals, significantly reducing computational time.

This strategy becomes particularly powerful when a closed-form expression for $\omega_{n,J}$ is available. For the {\em ALLS}, approximating the digammas by  logarithms, \eqref{BTZALLS} gives \cite{Burman1}
\beq
\omega_{n,J} \approx 
\frac{n\pi r_h}
{L^2\left[
\log\!\left(\frac{L}{\alpha \ell_p}\right)
-
\log\!\left(\frac{J L}{r_h}\right)
\right]},
\label{sliverspec}
\eeq
\begin{figure}[h]
       \centering
       \includegraphics[width=0.6\linewidth]{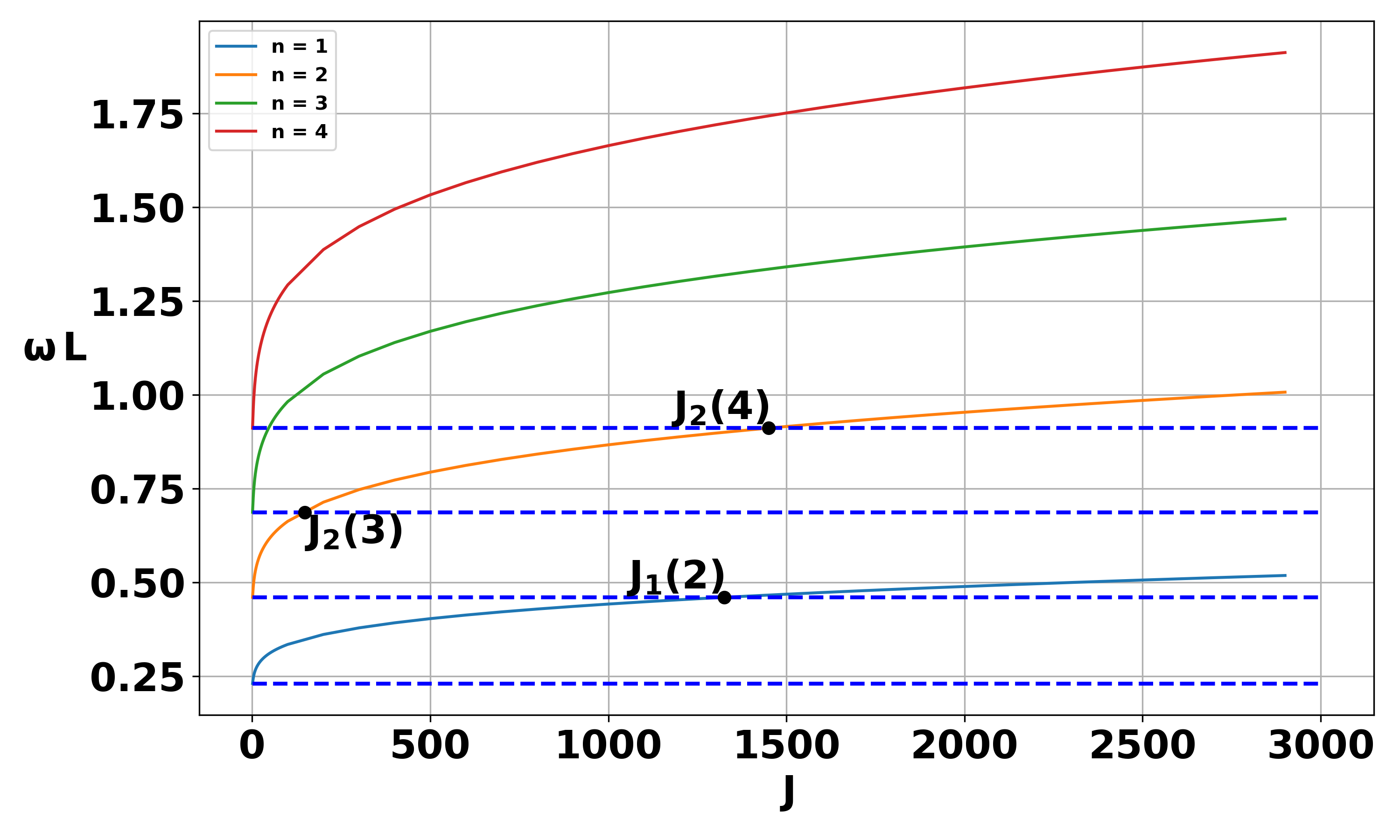}
       \caption{The intersection of the $n^{\text{th}}$ spectral line with $m\omega_0$ determines $J_{n}(m)$.}
       \label{slivercalc}
\end{figure}
where the term $\log\!\left(\frac{\omega L^2}{r_h}\right)$ has been neglected, as it is sub-leading in the thermodynamic regime (see Section~\ref{sec:BTZ}). Defining $J_n(m)$ through the inversion condition (see Fig. \ref{slivercalc})
\beq
\omega_{n,J_n(m)} = m\,\omega_0,
\label{Jnmdefinition}
\eeq
the number of $J$-modes for fixed $n$ whose frequencies lie in the interval $[m\omega_0,(m+1)\omega_0]$ is
\beq
\Delta J_n(m) = J_n(m+1) - J_n(m).
\eeq
Summing over all contributing $n$ gives the total degeneracy in that bin,
\beq
g = \sum_{n=1}^{m} \bigl[J_n(m+1) - J_n(m)\bigr],
\eeq
which can be expressed in closed form for {\em ALLS} after solving \ref{Jnmdefinition} for $J_n(m)$ as
\beq
g(J_{\rm max},m)
=
(J_{\rm max}-1)
\left(
\frac{1}{J_{\rm max}^{\frac{1}{m+1}} - 1}
-
\frac{1}{J_{\rm max}^{\frac{1}{m}} - 1}
\right),
\qquad
J_{\rm max} = \frac{r_h}{\alpha \ell_p}.
\eeq

The second-quantized partition function then reduces to
\beq
\log Z(\beta)
\approx
-2 \sum_{m=1}^{\infty}
g(J_{\rm max},m)
\log\!\left(1 - e^{-\beta m \omega_0}\right),
\label{sliverpartition}
\eeq
where the factor of $2$ accounts for the $J<0$ modes. The corresponding energy and entropy are
\beq
U \approx
2\sum_{m=1}^{\infty}
g(J_{\rm max},m)
\frac{m \omega_0}{e^{\beta m \omega_0} - 1},
\label{sliverenergy}
\eeq
\beq
S \approx
2\sum_{m=1}^{\infty}
g(J_{\rm max},m)
\left(
\frac{\beta m \omega_0}{e^{\beta m \omega_0} - 1}
-
\log\!\left(1 - e^{-\beta m \omega_0}\right)
\right).
\label{sliverentropy}
\eeq
Using $g(J_{\rm max},m)$ as defined above provides an upper bound, while replacing it by $g(J_{\rm max},m-1)$ in the above expressions yields the corresponding lower bound.
\section{Comments on 't Hooft's Calculation } \label{sec:tHooft}

In this appendix, we review 't Hooft's semi-classical calculation \cite{tHooft} of black hole entropy in the two simplest backgrounds: BTZ and Schwarzschild. His original analysis used the hard-wall cutoff for Schwarzschild. Here we extend the discussion to the BTZ case as well, and will also consider the holographic anchoring in both cases.

Our goal is to demonstrate two points. First, we observe that the real utility of the semi-classical approximation is that it effectively provides a {\em kinematic} cutoff for the spectrum, similar to the role played by $J_{\rm cut}$ in the degenerate spectrum approximation \cite{Burman1, Burman2}. In other words, the semi-classical approximation ``pushes down'' the modes and makes them ``more'' degenerate. This leads to an important feature {\em not} shared by the exact modes (but {\em is} a feature the degenerate modes): for sufficiently large $J$, the semi-classical modes go under the potential barrier. In this sense, the semi-classical approximation can be viewed as an alternative mechanism that effectively imposes a cutoff in $J$.

The second observation is that 't Hooft's calculation {\em also} implicitly contains the little hierarchy we have emphasized in this work. Although the semi-classical approximation makes the modes more degenerate than they actually are, the resulting spectrum is still not flat in $J$ -- this means that the hierarchy is present, but not as striking as it is for the exact spectrum. We believe this to be the reason why it was not noted in the literature previously. 

It is also worth noting that 't Hooft's semi-classical approximation with a horizon-anchored cut-off leads to a mismatch between the energy in the semi-classical normal modes and the mass of the black hole, even after we accept the hierarchy and match the entropies \cite{tHooft}\footnote{For the BTZ case (and not Schwarzschild), the semi-classical normal modes of the boundary-anchored case do not have this mismatch, but the little hierarchy is still present.}.

\subsection{BTZ}

For the massless case, the Bohr--Sommerfeld quantization condition in BTZ can be written as \cite{Solodukhin}
\beq \label{bohrsommquant}
n(\omega,J) = \frac{1}{\pi} \bigintsss_{r_\epsilon}^{r_{\omega,J}}\frac{\dd{r}}{f(r)}\sqrt{\omega^2 - \frac{J^2}{r^2}f(r)} .
\eeq
The integral is real only in the region where the expression inside the square root is positive. This condition determines the integration range of $r$, from the brick-wall radius $r_\epsilon$ up to the point where the integrand vanishes for given $\omega$ and $J$. This fixes $r_{\omega,J}$ to be
\beq
r_{\omega,J} = \frac{J r_h}{\sqrt{J^2 - \omega^2 L^2}} .
\eeq

The total number of modes with energy $\omega$ is
\beq \label{nomega}
n(\omega) = \frac{1}{\pi} \bigintsss_{r_\epsilon}^{r_{\omega,J}}\frac{\dd{r}}{f(r)} 
\bigintsss_0^{J_{\rm max}} \dd{J} 
\sqrt{\omega^2 - \frac{J^2}{r^2}f(r)},
\eeq
where positivity of the square root fixes
\beq \label{JmaxBTZ}
J_{\rm max} = \frac{\omega r}{\sqrt{f(r)}},
\eeq
for given $r$ and $\omega$. Performing the $J$ integral followed by the $r$ integral (both indefinite integrals can be done explicitly) and focusing on the lower limit of the latter, following 't Hooft, we obtain the leading-order expression
\beq \label{BTZnomegasoft}
n(\omega) \approx \frac{\omega^2 L^3}{4\sqrt{r^{2}_\epsilon - r^{2}_h}}
= \frac{\omega L^2}{4 \alpha \ell_p},
\eeq
where \eqref{boundarydef} has been used to obtain the second equality -- in other words, we are working with the boundary-anchored wall. (We will repeat the calculation also for the horizon-anchored case.)

We now consider the canonical ensemble of a scalar field at inverse temperature $\beta$. Each one-particle state can be occupied by any integer number of quanta, which leads to the second-quantized partition function. The free energy is therefore
\beq \label{FBTZsoft}
F = \frac{1}{\beta} \bigintsss_0^\infty \dd{\omega} 
\frac{\dd {n(\omega)}}{\dd{\omega}} 
\log(1 - e^{-\beta \omega})
= -\frac{\pi^2 L^2}{24 \alpha \beta^2 \ell_p}.
\eeq
From this we obtain the entropy ($S$) and average energy ($U$):
\begin{align} \label{SUBTZsoft}
S = \beta^2 \partial_\beta F 
= \frac{1}{12 \alpha} \left(\frac{\pi r_h}{2 \ell_p}\right), 
\qquad
U = F + \frac{S}{\beta} 
= \frac{1}{12 \alpha} \left(\frac{r^{2}_h}{8 \ell_p L^2}\right).
\end{align}
These expressions are evaluated at the Hawking temperature for BTZ,
$\beta_H = \frac{2 \pi L^2}{r_h}$.

An interesting fact is that the $\alpha$-dependent constant is identical for both the entropy and the average energy. This is a specific feature of the holographic anchoring (and only for BTZ), which no longer holds when we go over to the horizon-anchored brick wall. Matching our result precisely to the geometric entropy,
$\frac{A}{4 G^{(3)}_N} = \frac{2 \pi r_h}{4 \ell_p}$,
requires choosing
\beq
\alpha = \frac{1}{12} = 0.0833\ldots ,
\eeq
which simultaneously sets the average energy equal to the physical mass,
$U = M_{BH} = \frac{r^{2}_h}{8 \ell_p L^2}$. In other words, with the semi-classical approximation for the modes and the holographic anchoring, the little hierarchy in the BTZ case is quite mild {\em and} the mass and entropy can both be made to match. It is not clear to us if this is a hint of something deeper: but with the exact spectrum the hierarchy is worse and the mismatch is also present, so we will not dwell on this further. Note also that when the spectrum is degenerate, there is no hierarchy and no coefficient mismatch \cite{Burman1, Burman2}.

We now compute the hard-wall cutoff case. The above calculation is insensitive to the kind of anchoring until the evaluation of $n(\omega)$. Using the first equality of \eqref{BTZnomegasoft} for $\epsilon \ll r_h$, we obtain
\beq
n_{h} (\omega) \approx \frac{\omega^2 L^3}{4\sqrt{2 r_h \epsilon}}
= \frac{\omega^2 L^4}{4 \alpha_h r_h \ell_p},
\eeq
where \eqref{slp} has been used in writing the final expression. This gives
\begin{align}
F_h &= -\frac{\zeta(3) L^4}{2 \alpha_h \beta^3 r_h \ell_p}, \\
S_h &= \frac{3 \zeta(3)}{4 \pi^3 \alpha_h} \left(\frac{\pi r_h}{2\ell_p}\right), \\
U_h &= \frac{\zeta(3)}{\pi^3 \alpha_h} \left(\frac{r^{2}_h}{8 \ell_p L^2}\right).
\end{align}
A notable difference from the previous case is that the $\alpha$-dependent constants for entropy and energy are no longer identical. Matching the entropy to the geometric value requires
\beq
\alpha_h = \frac{3 \zeta(3)}{4 \pi^3} = 0.0290\ldots ,
\eeq
which implies $U_h = \frac{4}{3} M_{BH}$.

We can also obtain an approximate expression for the spectrum by explicitly performing the $r$ integral in \eqref{bohrsommquant}. Evaluating the result at the lower limit $r_\epsilon$ (since the upper limit does not contribute at leading order), we obtain the following approximate relation for $\omega_{n,J}$:
\beq \label{thooftspectrum}
\frac{\omega L^2}{r_h}\approx 
\frac{ n\pi}
{\log\left(\frac{2L}{\alpha \ell_{p}}\right)
- \log\left(\frac{JL}{r_h}\right)
+ \log \left(\frac{\omega L^2}{r_h}\right)} .
\eeq
We will call this the 't Hooftian semi-classical spectrum.
\begin{figure}[h]
       \centering
       \includegraphics[width=0.6\linewidth]{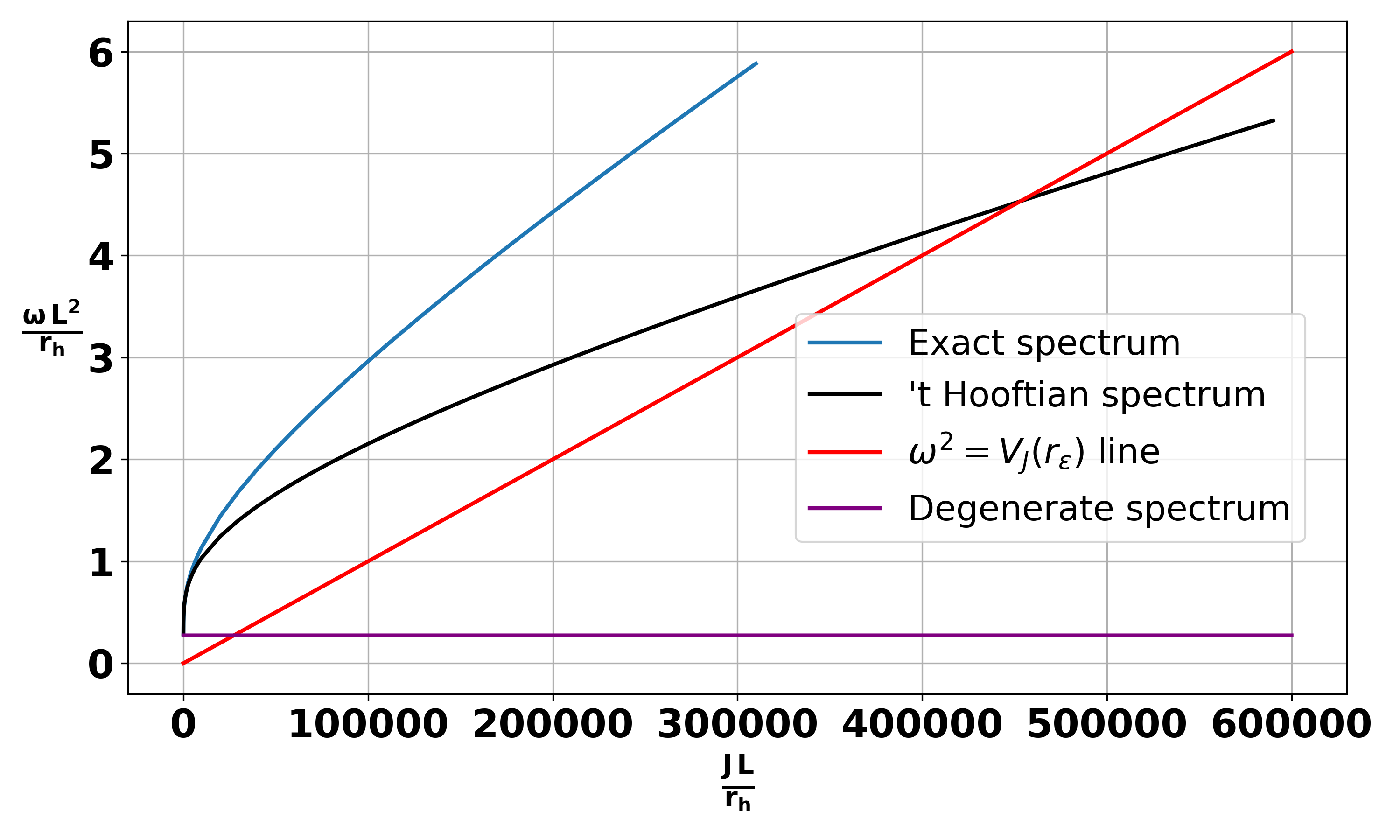}
       \caption{Comparison between the hard wall exact spectrum and the ’t Hooftian semi-classical spectrum (\ref{thooftspectrum}) for $\widetilde{\ell}=10^{-5}$, $n=1$ and arbitrary $r_h/L$. The red line represents $\omega^2 = V_J(r_{\epsilon})$: energies above this line are quantum mechanically allowed. Note that the growth of $\omega$ is hierarchically slow with $J$ for all three of the spectra, compared to $n$. This is the reason why the curves look continuous, and this is the sense in which the spectrum is approximately degenerate in $J$. But note that the ranges of $J$'s present in the different cases are quite different -- this is one way to think about the little hierarchy.}
       \label{Exactvsthooftspectrum}
\end{figure}

A comparison of this spectrum with the exact spectrum is shown in Fig.~\ref{Exactvsthooftspectrum}. One clearly observes that the semiclassical spectrum is ``pushed down'' relative to the exact one. This explains why the value of $\alpha$ obtained in the WKB analysis is slightly larger (better!) than the value obtained from the exact spectrum, since pushing the spectrum downward allows energies to contribute more.

This behavior is a natural feature of spectra obtained from Bohr-Sommerfeld quantization. The determination of the semiclassical spectrum relies on the intersection $\omega^2 = V_J(r)$, and therefore it is natural that the spectrum gets pushed down compared to the exact spectrum. Quantum mechanically, however, the exact spectrum does not meet this curve and remains above it. Consequently, semiclassical analyses generally give a slightly better value of the hierarchy parameter $\alpha$ compared to the value obtained from a full statistical-mechanical calculation using the exact spectrum.

\subsection{Schwarzschild}

The Schwarzschild background is spherically symmetric, so the metric is given by a blackening factor $f(r) = 1 - \frac{r_h}{r}$ and the angular momentum controlled by $l(l+1)$. 
Thus, the total number of modes with energy $\omega$ in Bohr-Sommerfeld quantization becomes
\beq \label{nomegal}
n(\omega) =
\frac{1}{\pi} 
\bigintsss_{r_\epsilon}^{r_{\omega,l}}
\frac{\dd{r}}{f(r)}
\bigintsss_0^{l_{\rm max}}
\dd{l}(2l+1)
\sqrt{\omega^2 - \frac{l(l+1)}{r^2}f(r)},
\eeq
where the degeneracy factor $2l+1$ arises from spherical symmetry. The reality condition fixes
\beq \label{lmaxSch}
l_{\rm max}(l_{\rm max}+1) = \frac{\omega^2 r^2}{f(r)}.
\eeq
Following the same steps as in the BTZ case, we obtain the leading-order expression
\beq
n(\omega) \approx 
\frac{2 \omega^3 r^{3}_h r_\epsilon}{3 \pi (r_\epsilon - r_h)}
+ \mathcal O\!\left(\log\!\left(\frac{r_\epsilon}{r_h} - 1\right)\right)
=
\frac{2 \omega r^{3}_h}{3 \pi \alpha^2 l^2_p}
+ \mathcal O\!\left(\log\!\left(\frac{\omega^2 \alpha^2 l^2_p}{1 - \omega^2 \alpha^2 l^2_p}\right)\right),
\eeq
Using the Schwarzschild version of the Boundary-anchored brickwall \eqref{boundarydef},
\beq
\frac{r_\epsilon}{r_h} = \frac{1}{1 - \omega^2 \alpha^2 l^{2}_p},
\eeq
we obtain the leading-order thermodynamic quantities
\begin{align} \label{SUSchsoft}
F &= -\frac{\pi r^{3}_h}{9 \alpha^2 \beta^2 l^{2}_p}, \\
S &= \frac{1}{18 \pi \alpha^2}\left(\frac{\pi r^{2}_h}{l^{2}_p}\right), \\
U &= \frac{1}{72 \pi \alpha^2} \left(\frac{r_h}{2 l^{2}_p}\right),
\end{align}

Unlike the BTZ case, the $\alpha$-dependent constants in entropy and energy differ (by a factor of $4$). Matching the entropy to the geometric value,
$\frac{A}{4 G^{(4)}_N} = \frac{4 \pi r_h^2}{4 l_p^2}$,
requires
\beq
\alpha = \frac{1}{\sqrt{18 \pi}} = 0.1329\ldots ,
\eeq
which implies $U = \frac{1}{4} M_{\rm BH}$. This is again a mild hierarchy: we will see that the hierarchy is again slightly worse in the horizon-anchored case. In both cases, the energy/entropy mismatch is still present.

Repeating the calculation for the hard-wall cutoff and using $\epsilon \ll r_h$ gives
\beq
n_h(\omega) \approx 
\frac{2 \omega^3 r^{4}_h}{3 \pi \epsilon}
=
\frac{8 \omega^3 r^{5}_h}{3 \pi \alpha^2 l^2_p},
\eeq
Using
\beq
s \approx 2 \sqrt{r_h \epsilon} \equiv \alpha \ell_p
\quad \Rightarrow \quad
\epsilon = \frac{\alpha^2 l^{2}_p}{4 r_h},
\eeq
we obtain
\begin{align}
F_h &= -\frac{8 \pi^3 r^{5}_h}{45 \beta^4 l^2_p}, \\
S_h &= \frac{1}{90 \pi \alpha^2_h} \left(\frac{\pi r^{2}_h}{l^{2}_p}\right), \\
U_h &= \frac{1}{240 \pi \alpha^2_h} 
\left(\frac{r_h}{2 l^{2}_p}\right),
\end{align}
Matching the entropy to the geometric value gives
\beq
\alpha_h = \frac{1}{\sqrt{90 \pi}} = 0.0594\ldots ,
\eeq
which implies
\[
U_h = \frac{3}{8} M_{BH},
\]
reproducing 't Hooft's original result \cite{tHooft}.

As pointed out in the Introduction, the hierarchy parameter $\alpha$ is encoded in the relation between the Planck length $l_p$ and the coordinate cutoff $\epsilon$. The broad lesson of these 't Hooft-inspired calculations is that the hierarchy and mismatch are both present in the semi-classical approximation, but both are somewhat ameliorated because the semi-classical approximate modes are ``more degenerate" than the actual modes.

\end{document}